\documentclass[preprint,showpacs,preprintnumbers,amsmath,superscriptaddress,prb]{revtex4}

\usepackage{amsfonts}
\usepackage{graphicx}

\newcommand{\tB}{\ensuremath t_\mathrm{B}}
\newcommand{\eigE}{\ensuremath \mathcal{E}}
\newcommand{\eigEp}{\ensuremath \mathcal{E}'}
\newcommand{\Knum}{\ensuremath \mathcal{K}}
\newcommand{\dd}{\ensuremath\mathrm{d}}
\newcommand{\tGE}{\ensuremath t_\mathrm{GE}}
\newcommand{\mmax}{\ensuremath m_\mathrm{max}}
\newcommand{\tsw}{\ensuremath t_\mathrm{sw}}

\newcommand{\Nb}{\ensuremath N_\mathrm{b}}
\newcommand{\NL}{\ensuremath N_\mathrm{L}}
\newcommand{\Nel}{\ensuremath N_\mathrm{el}}

\newcommand{\HleadL}{\ensuremath \hat{H}^\mathrm{lead}_L}
\newcommand{\HLeads}{\ensuremath \hat{H}^\mathrm{Leads}}
\newcommand{\HFD}{\ensuremath \hat{H}^\mathrm{FD}}
\newcommand{\HFDLC}{\ensuremath \hat{H}^\mathrm{FDLC}}
\newcommand{\HBias}{\ensuremath \hat{H}^\mathrm{Bias}}

\begin{document}
\title{Wavepacket representation of leads for efficient simulations of time-dependent electronic transport}
\author{Martin Kon\^{o}pka}
\affiliation{
Department of Physics,\\
Institute of Nuclear and Physical Engineering,\\
Faculty of Electrical Engineering and Information Technology,\\
Slovak University of Technology in Bratislava,\\
Ilkovi\v{c}ova 3, 812~19 Bratislava, Slovakia
}
\author{Peter Bokes}
\affiliation{
Department of Physics,\\
Institute of Nuclear and Physical Engineering,\\
Faculty of Electrical Engineering and Information Technology,\\
Slovak University of Technology in Bratislava,\\
Ilkovi\v{c}ova 3, 812~19 Bratislava, Slovakia
}
\date{\today}
\begin{abstract}
We present theoretical foundations and numerical demonstration of an efficient method
for performing time-dependent many-electron simulations for electronic transport.
The method employs the concept of stroboscopic wavepacket basis
for the description of electrons' dynamics in the semi-infinite leads.
The rest of the system can be treated using common propagation schemes for finite electronic systems.
We use the implementation of our method to study the time-dependent current response in armchair graphene
nano-ribbons (AGNRs) with sizes up to 800 atoms described within tight-binding approximation.
The character of the time-dependent current is studied for different magnitudes of the bias voltage,
variable width and length of AGNRs, different positions of the current measurement, and for full and reduced coupling
of the AGNRs to the electrodes.
\end{abstract}
%
\maketitle
%
%
\section{\label{sec:intro}Introduction}
%
%
Continual miniaturization of electronic technology is coming to its ultimate limit 
where a single circuit element might consist of several tens or hundreds of atoms only. 
Molecular nanotransistors and nanotransistors based on graphene nano-ribbons are examples of this development. 
To understand temporal behavior of these devices, e.g. their 
switching or operation at GHz to THz regimes, one has to use time-dependent quantum-mechanical 
model of open electronic systems. It is also desirable that the model captures 
the chemical character of involved constituents, for which the first-principles methods 
are suitable. The combination of these two requirements represents a challenge for numerical simulations.

The formulation of the time-dependent quantum-mechanical electronic transport in nanojunctions 
has been put forward by Jauho, Meir and Wingreen~\cite{Jauho1994} within the framework of non-equilibrium 
Green's functions (NEGF), for a model that consists of non-interacting leads and possibly interacting finite 
central part. Its applications to transport problems in mesoscopic and nanoscopic transport are
immense, such as photon assisted tunneling~\cite{Stafford1996} or Kondo physics~\cite{Nordlander1999},
but the models used typically consisted of few one-particle sites only. The electron-electron correlation,
if accounted for at all, is usually included through a model Hubbard term.
To go beyond these models, one generally needs 
larger one-particle basis, e.g. several atomic orbitals per atom, and a theory that in spite of this increase 
of the one-particle basis is capable to describe electron-electron interaction to a satisfactory level.

The time-dependent density-functional theory (TDDFT) or the time-dependent current-density 
functional theory are most likely to be practically useful for systems with few tens of atoms and hundreds 
of one-particle basis functions in the central region. Several authors have discussed the applicability 
of the TDDFT for quantum transport either in the framework of large but finite systems~\cite{DiVentra04b} 
or using non-partitioning approach~\cite{Stefanucci04}. Problems of the adiabatic approximation 
for the exchange-correlation potential has been discussed~\cite{Sai2005,Koentopp2008,Mera2010,Kurth2010} and 
possible improvements explored using a comparison between the TDDFT and many-body 
techniques~\cite{Myohanen2009,Schmitteckert2008,Ramsden2012,Kurth2013}, but the identification of a suitable 
exchange-correlation potential is still an open and very important problem.

Several authors proposed methodologies to propagate the NEGF equations in time numerically,
assuming that some sort of self-consistent time-dependent exchange-correlation potential is available.
Zhu \textit{et al.}~\cite{Zhu2005} made use of finite correlation time of one-electron Green's functions to 
propagate the equations of motion for the Green's function. 
This time-domain decomposition technique has been later used to study the transport under various 
time-dependent biases through hydrogen molecule coupled to hydrogen 1D chains, which demonstrated 
the applicability of this method to more realistic models~\cite{Ke2010}.
Significant improvement in numerical cost can be achieved using complex absorbing boundary conditions.
Quality of this approximation has been recently tested and demonstrated for a nanojunction with a short carbon
chain bridging bulk Al electrodes~\cite{Zhang2013}. 
For small model systems the direct propagation of the NEGF equations in time has been implemented even beyond 
the TDDFT framework, namely using the many-body perturbation theory~\cite{Myohanen2009}.

Clearly, propagating the Green's functions is numerically more demanding than propagating time-dependent
wavefunctions. Kurth et. al.~\cite{Kurth2005} casted the time-dependent NEGF formulation into 
the effective time-dependent Schr\"{o}dinger equation, while preserving the coherence of the electrons 
transported from and to the leads. 
Approximations in the treatment of the lead's self-energy using the above mentioned
complex absorbing potential has been used to compare the long-time steady state and stationary 
non-equilibrium calculation of transport properties of benzene-dithiol molecule anchored to two semi-infinite
gold chains~\cite{Varga2011}.
Numerically efficient approach based on time-dependent scattering theory 
has been recently presented by Gaury \textit{et al.}~\cite{Gaury2013},
based on time-dependent scattering methods, which apart from demonstrative calculations contains 
detailed comparison between various presently considered methods to address time-dependent
electronic transport.

An alternative way for the description of fully coherent dynamics of electrons in open systems is
offered by the stroboscopic wavepacket basis~\cite{Bokes_PRL}. Later we have reformulated this method 
using time-dependent basis functions~\cite{Bokes_PCCP}. The use of time-dependent basis 
for quantum transport, but not in the context of stroboscopic wavepackets, has been also independently 
suggested by Varga~\cite{Varga2012}. Recently we have presented an implementation of the stroboscopic 
wavepacket approach (SWPA), that employs this basis within the whole system for treatment 
of electronic quantum transport through atomistic models of nanojunctions~\cite{our_EPJB}.
The method exploits several features of the stroboscopic wave packets:
(i)~their partially localized nature,
(ii)~their mutual orthonormality and completeness,
(iii)~their unitary propagation such that the basis set evolves in
time.
Using these properties within the SWPA the open boundary conditions are incorporated in a straightforward
way with the number of explicitly included electrons being variable.
However, the not-so-well localized wavepackets caused serious convergence difficulties 
in computed quantities like electric currents, especially at higher biases~\cite{our_EPJB}.

In the present work we generalize the above method using a mixed basis set consisting
of stroboscopic wave packets within the semi-infinite leads of the system and 
localized atomic orbitals in the finite central part of the nanojunction.
As we will demonstrate in our paper, this generalization removes the convergence difficulties 
already at rather small basis set sizes which makes the method particularly suitable for 
first-principles time-dependent transport simulations.
In addition to the increased numerical efficiency, the generalized method enables 
us to study systems with arbitrary number of leads, which can be conveniently used to describe 
geometrically wide leads or multiterminal nanodevices.
To keep the presentation and the first implementation simple and comparable to other 
proposed schemes, we employ only the tight-binding (TB) description for electrons in both the leads 
as well as the central region.

The rest of the paper is divided into two main parts. In Sec.~\ref{sec:method} we give the theoretical 
foundations for the generalized stroboscopic wavepacket approach.
In the second part, in Sec.~\ref{sec:graphene}, we demonstrate the power of the method by performing
extensive study of time-dependent transport in graphene nano-ribbons composed of hundreds of atoms.
We note that in the appendix section~\ref{sec:rings} we demonstrate the improvement of the generalized 
method over our former implementation by applying the new method to the simulation of electron 
currents in quantum rings, the model which has been addressed in the past~\cite{our_EPJB}.
Finally, in Sec.~\ref{sec:conclu} we discuss the results and make conclusions.
%
%
\section{\label{sec:method}The method}
%
%
\subsection{\label{ssec:model} Model of multi-terminal nanojunction}
\begin{figure}[!thb]
\centerline{\includegraphics[width=0.40\textwidth]{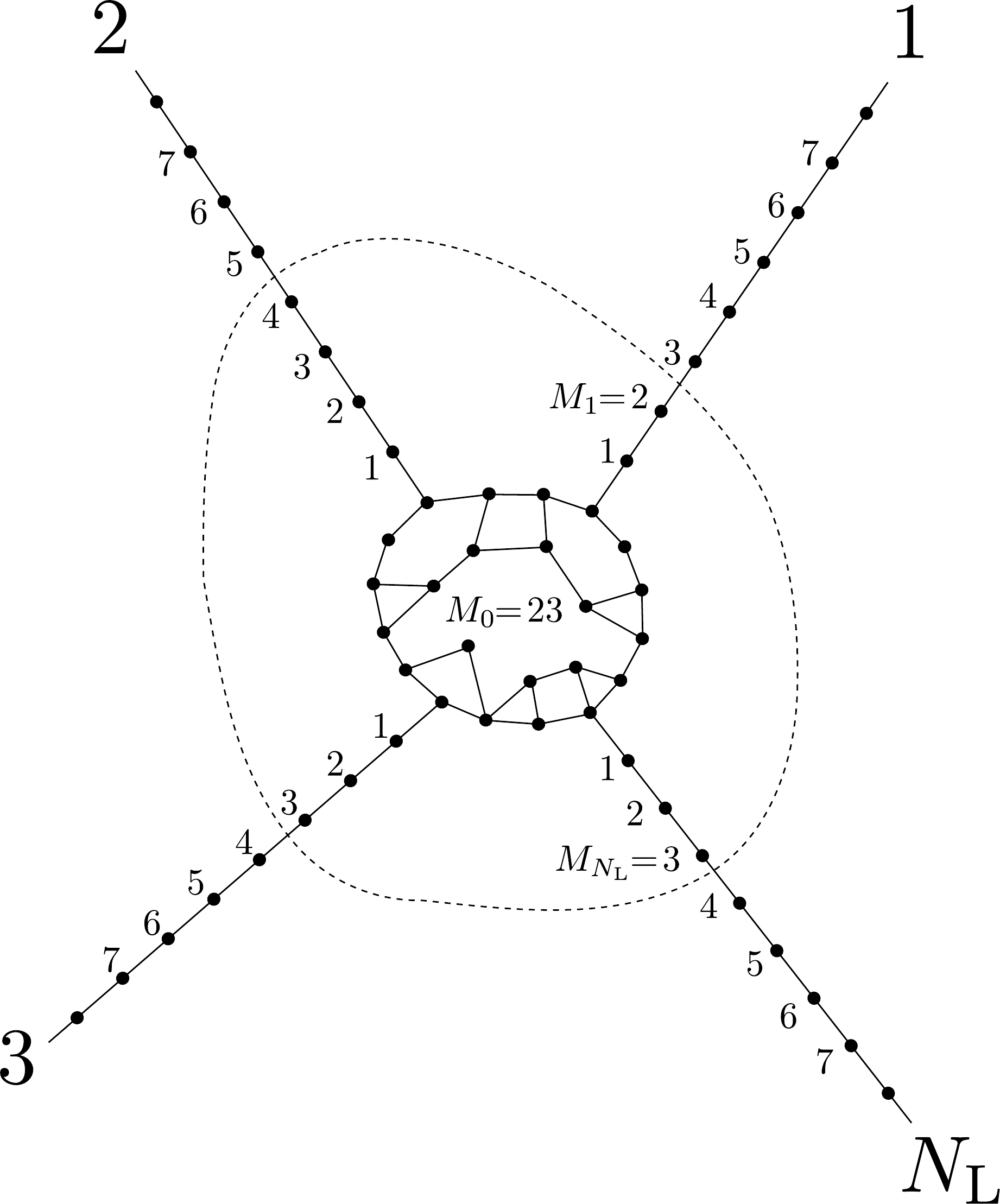}}
\caption{The scheme of the multi-lead device.
In this example we use $\NL = 4$.
The dashed line divides the formal device from the rest of the entire system.
The numbers $M_L$ with $L \in \{1, \dots, \NL\}$ set the outer boundaries of the formal device.
The physical device contains $M_0 = 23$ atoms in this example.
See main text for more description.}
\label{fig:theory_junction}
\end{figure}

%
%
Our aim is to describe a multi-terminal nanojunction that consists of a central part, 
which we will refer to as the \emph{physical} device $\mathbb{D}_0$, and of $\NL$ semi-infinite 
electrodes.
Alternatively, we will also partition the same total system into the \emph{formal} device $\mathbb{D}$ that consists of 
the physical device together with finite segments of the electrodes, and 
the remaining portions of the electrodes which we will call \emph{the leads}.
These two possible partitions of the total system are illustrated in Fig.~\ref{fig:theory_junction},
%
%
where the formal device is inside the dashed-line surrounded region.

The division into formal device and leads is motivated by the expectation that 
sufficiently far from the physical device the electronic structure in the electrodes
will attain semi-periodic and nearly equilibrium character. Hence, the lead is such a semi-infinite part 
of the electrode that is already in this idealized state. Dynamics of electrons in the leads
will be efficiently described within the stroboscopic wavepacket representation, as will be given
in Sec.~\ref{ssec:strobo}.

Our present numerical implementation uses tight-binding description of the electronic structure of the nanojunction
and the below given exposition is done within this framework.\footnote{More generally, all the formulas 
can be directly written also for any time-dependent self-consistent method within use with localized basis, e.g. 
the time-dependent density functional theory.}
Using the partitioning into the formal device and the leads, the total Hamiltonian $\hat{H}(t)$ is split
into four parts:
\begin{equation}
\hat{H}(t) = \HLeads + \HFD(t) + \HFDLC(t) + \HBias(t) \ .
\label{eq:H}
\end{equation}
$\HLeads = \sum_{L=1}^{\NL} \HleadL$ is the time-independent Hamiltonian of the leads in equilibrium, with
\begin{equation}
\HleadL
=
\sum_{l=M_L+1}^\infty  \epsilon a^\dag_{L,l}  a_{L,l}
+
\sum_{l=M_L+1}^\infty \tB (a^\dag_{L,l+1}  a_{L,l}  + a^\dag_{L,l} a_{L,l+1})
\ ,
\label{eq:HleadL}
\end{equation}
where $\epsilon$ is the orbital (on-site) energy and $\tB<0$ 
is the inter-site hopping matrix element. The operators $a^\dag_{L,l}$ and $a_{L,l}$
create and destroy an electron in leads' orbitals $|L,l\rangle$.
The pair of the indices $L=1,\dots,\NL$ and $l = M_L+1, M_L+2, \dots$ denote lattice site $l$ in lead $L$.
Orbitals $|L,l\rangle$ are presumably orthonormal and form a complete basis for the leads.

$\HBias(t)$ represents applied electric biases in the leads, which depend on time but not on the 
site index within the lead,
\begin{equation}
\HBias(t)
=
\sum_{L=1}^{\NL} \, \sum_{l = M_L+1}^\infty e U_L(t) \, a^\dag_{L,l} a_{L,l}
\ .
\label{eq:HBias}
\end{equation}
$e$ is the unit charge, $U_L(t)$ is the (generally time-dependent)
bias applied to lead $L$, and the index $l$ labels individual orbitals in the lead.

$\HFD(t)$ describes the time-dependent dynamics of electrons in the formal device without coupling to the leads.
It can be written in the form
\begin{equation}
\HFD(t)
=
\sum_{L=0}^{\NL} \sum_{l=1}^{M_L} \epsilon_{L,l}(t) \, a^\dag_{L,l} a_{L,l}
\, + 
\sum_{L,L'=0}^{\NL} \; \sum_{l,l'=1}^{M_L} t_{L,l;L',l'}(t) \, a^\dag_{L,l} a_{L',l'}
\ .
\label{eq:HFDgen}
\end{equation}
The relevant orbitals $|L,l\rangle$ belong either to the electrodes 
with $L \in \{1,..,N_L\}$ and the site indices $l \in \{1, 2, \dots, M_L\}$ or to the physical device
indicated with the index $L = 0$.
$\HFD(t)$ therefore includes also the degrees of freedom in the finite pieces of the electrodes,
possible effects of a gate potential, and the form of the time-dependent bias voltages within the formal device. 
$\epsilon_{L,l}(t)$ and $t_{L,l;L',l'}(t)$ are the corresponding diagonal and off-diagonal matrix elements of $\HFD(t)$,
respectively.
In the present work we use time-independent hopping matrix elements $t_{L,l;L',l'}(t)$
which are either zeros (no direct couplings between individual electrodes), $\tB$ (between nearest-neighbor atoms),
or optionally a reduced coupling $0.25\,\tB$ between the electrodes and the physical device.

The bias within the finite pieces of the electrodes has the same form as in $\HBias(t)$.
Within the physical device the bias contributions are denoted as $e U_{0,l}(t)$ and can be set arbitrarily.
The bias contributions within the entire formal device are added to the ground-state atomic on-site energies $\epsilon$,
i.e. they are included in time-dependent on-site energies $\epsilon_{L,l}(t)$ introduced in eq.~(\ref{eq:HFDgen}).
More rigorous approach would be to compute these quantities by taking into
account the Coulomb interactions in the system.
Since our present formulation of the method works with independent electrons,
we instead use a prescribed value of the bias effect on the physical device.
One possibility for this prescription is to use an average of the biases applied to the individual leads,
$U_{0,l}(t) = U_{0}(t) = 1/\NL \, \sum_{L=1}^{\NL} U_L(t)$,
identical for all sites of the physical device. 
This models a weak symmetric coupling regime with delocalized (well conducting) states
that are able to screening out the applied bias within the physical device.
The second possibility,
used in the present work (the exception being Appendix~\ref{sec:rings})
assumes a linear variation of $U_{0,l}(t)$ between the source lead and the drain
where the applied bias can not be screened out in the central device (see section~\ref{sec:graphene} dealing
with the graphene nano-ribbons).

Finally, the coupling between the finite pieces of the electrodes and the leads
is described by the term $\HFDLC(t)$ (formal-device-to-leads coupling).
Although this operator could in generally be time-dependent,
here we choose the simple tight-binding time-independent form
\begin{equation}
\HFDLC
=
\sum_{L=1}^{\NL} \tB (a^\dag_{L,M_L+1} a_{L,M_L} + a^\dag_{L,M_L} a_{L,M_L+1})
\label{eq:HFDLC}
\end{equation}
Implications of this form as well as of the other terms of Hamiltonian~(\ref{eq:H}) are described in Sec.~\ref{ssec:SchE}.
%
%
\subsection{\label{ssec:strobo}Multiple semi-infinite leads and the stroboscopic wavepackets}
%
%
The dynamics of electrons within the leads is described using the stroboscopic wavepackets 
representation~\cite{Bokes_PRL,Bokes_PCCP,our_EPJB}. 
The unique feature of the stroboscopic wavepacket
basis set is that time-propagation of each basis function (a wavepacket) in finite time step $\tau$,
governed by a suitable Hamiltonian, results in its mapping into another basis function of the basis. 
Hence, part of the dynamics of electrons is already built into the basis set, and the whole time-dependent 
basis set maps onto itself after the time $\tau$.  
For a detailed description of this representation we refer the reader to its first exposition
using only stationary basis set and real-space formulation~\cite{Bokes_PRL}, its extension 
for time-dependent basis~\cite{Bokes_PCCP} and finally its implementation for an infinite one-dimensional tight-binding
model~\cite{our_EPJB}.

In contrast with the previous work, here it is implemented 
for $\NL$ disconnected semi-infinite leads in the form of one-dimensional tight-binding chains. 
In the following we summarize the construction of the wavepackets, giving details only for the 
new results that are specific for the semi-infinite character of the leads.

The Hamiltonian $\HleadL$ of each lead (eq.~\ref{eq:HleadL})
has a continuous spectrum of non-degenerate eigenstates,
\begin{equation}
\HleadL |L,\eigE\rangle = \eigE |L,\eigE\rangle\ .
\label{eq:eigsysL}
\end{equation}
The eigenstates are orthogonal, their products being
\begin{equation}
\langle L',\eigEp|L,\eigE\rangle = \delta_{L,L'}\, \delta(\eigE-\eigEp)\ .
\end{equation}
In the stroboscopic-wavepacket method, the eigenenergy spectrum is arbitrarily split into $\Nb$ \emph{bands},
\begin{equation}
\eigE \in [\eigE_0,\eigE_1] \cup [\eigE_1,\eigE_2] \cup \dots \cup [\eigE_{\Nb-1},\eigE_{\Nb}] \ ,
\label{eq:bands}
\end{equation}
and for each band one defines the temporal distance $\tau_n$ between two consecutive wavepackets of the basis:
\begin{equation}
\tau_n = \frac{2\pi\hbar}{\Delta\eigE_n}
\label{eq:tau_n}
\end{equation}
with
\begin{equation}
\Delta\eigE_n \equiv \eigE_n - \eigE_{n-1} \ . 
\end{equation}
The unitarily propagating stroboscopic basis set is then defined (see also~\cite{Bokes_PRL,Bokes_PCCP,our_EPJB})
as the set of orthonormal vectors
\begin{equation}
|L,n,m;t\rangle
=
\frac{1}{\sqrt{\Delta\eigE_n}} \int_{\eigE_{n-1}}^{\eigE_n} \exp\left[-\frac{i}{\hbar} \eigE (t+m\tau_n) \right]
|L,\eigE\rangle \, \dd\eigE \ ,
\label{eq:packet_def}
\end{equation}
where
the indices are running through the ranges $L \in \{1, \dots, \NL\}$,
$n \in \{1, \dots, \Nb\}$, and $m \in \mathbb{Z}$.
The wavepackets~(\ref{eq:packet_def}) form an orthogonal set that spans the same Hilbert space as the original 
eigenstates (\ref{eq:eigsysL}) and in this sense is complete.
The eigenstates $|L,\eigE\rangle$ of the semi-infinite lead $L$ can be expressed by the expansion
\begin{equation}
|L,\eigE\rangle
=
\sum_{l=M_L+1}^\infty \psi_{L,l} |L,l\rangle
\ ,
\end{equation}
where the coefficients $\psi_{L,l}$ are
\begin{equation}
\psi_{L,l}
=
\frac{1}{\sqrt{\pi |\tB| \sin\Knum}} \, \sin[\Knum (l-M_L)] \ , \ \ \ \ l \ge M_L+1
\end{equation}
and the wavenumber $\Knum>0$ is related to the eigenenergy of the state:
\begin{equation}
\eigE = \epsilon + 2 \tB \cos\Knum \ .
\end{equation}
Stroboscopic wavepackets~(\ref{eq:packet_def}) for the TB model under consideration can
be represented in the orbital basis set:\footnote{See also Appendix in Ref.~\onlinecite{our_EPJB} where similar
overlaps have been derived in the context of the original SWPA.}
\begin{eqnarray}
\nonumber
& &\langle L,l|L,n,m;t\rangle
=
2 \, \sqrt{\frac{|\tB|}{\pi\Delta\mathcal{E}_n}} \exp\left(-\frac{i}{\hbar} \epsilon t'\right)
\\
&\times& \int_{\Knum_{n-1}}^{\Knum_n} \sqrt{\sin\Knum} \, \exp\!\left[-\frac{i}{\hbar}(2\tB \cos\Knum) t'\right] \sin[\Knum (l-M_L)]
\, \dd\Knum
\label{eq:ovlp_reflect}
\end{eqnarray}
for $l \ge M_L+1$.
The symbol
$t' \equiv t + m \tau_n$
helps to keep the formula more compact. We evaluate integrals in overlaps~(\ref{eq:ovlp_reflect}) numerically.

The states (\ref{eq:ovlp_reflect}) with progressing time $t$ constitute wavepackets coming from sites with large $l$
towards $l=M_{L}$, and consequently reflecting back into the lead.
For this reason we refer to the sites with $l = M_L$ within the formal device as the \emph{mirror} sites.
%
%
\subsection{\label{ssec:SchE}The basis set and the Schr\"{o}dinger equation}
%
%
The mixed basis set used in our approach consists of the
unitarily propagating stroboscopic wavepackets~(\ref{eq:packet_def})
covering the leads 
and from a different set of basis functions which cover the formal device.
We will use a shorthand notation for the stroboscopic basis vectors within the leads 
using the composite index $o$,
\begin{equation}
|o;t\rangle \equiv |L, n, m; t\rangle \, ,
\label{eq:strobo_notations}
\end{equation}
and similarly for the basis functions in the formal device,
\begin{equation}
|u\rangle \equiv |L,l\rangle \ ,
\label{eq:orbital_notations}
\end{equation}
using the composite index $u$.
The stroboscopic vectors are mutually orthonormal.
We also assume that the vectors $|u\rangle$ are mutually orthonormal.
In addition we construct the basis set so that the orthogonality
\begin{equation}
\langle o;t|u\rangle = 0
\end{equation}
is satisfied for any pair of $|o;t\rangle$ and $|u\rangle$
and the resulting set of vectors
$\{|o;t\rangle, |u\rangle\}$ forms a complete and orthonormal basis.
In practice we will have to make cutoffs on the number of the stroboscopic
wave packets.
This is accomplishes by using a finite maximum index $m$,
denoted as $\mmax$, in the basis set~(\ref{eq:packet_def}).

The mixed basis set is used to solve the Schr\"{o}dinger equation (SchE)
\begin{equation}
i \hbar \frac{\dd}{\dd t} |\Psi(t)\rangle = \hat{H}(t) |\Psi(t)\rangle
\label{eq:SchE}
\end{equation}
for the entire system described by Hamiltonian~(\ref{eq:H}).

The state vector (of a single electron) is expressed in the above-defined
basis set:
\begin{equation}
|\Psi(t)\rangle = \sum_o \mathcal{A}_o(t) |o; t\rangle  +  \sum_u A_u(t) |u\rangle
\ .
\label{eq:state_vector}
\end{equation}
$\mathcal{A}_o(t)$ and $A_u(t)$ are the probability amplitudes to be determined
by solving the SchE.
State vector~(\ref{eq:state_vector}) is substituted 
into the SchE and subsequently the properties of the 
stroboscopic wavepackets and the localized orbitals are utilized.
First, the stroboscopic vectors $|o; t\rangle$ satisfy the equation
\begin{equation}
\frac{\dd}{\dd t} |o; t\rangle  =  -\frac{i}{\hbar} \HLeads |o; t\rangle \ ,
\label{eq:strobo_vector_evo}
\end{equation}
which results in elimination of two terms in the SchE expressed in the considered
basis set.
Another simplification comes from the fact that
\begin{equation}
\HLeads |u\rangle = 0 \ \ \textrm{for each} \ u \in \mathbb{D}
\end{equation}
because $\HLeads$ does not involve any operator belonging to the formal device.
Similarly we have
\begin{equation}
\HBias(t) |u\rangle = 0 \ \ \textrm{for each} \ u \in \mathbb{D}
\end{equation}
since, again, $\HBias(t)$ by its definition involves only degrees of freedom
from exterior of the formal device.
On the other side,
the formal-device Hamiltonian $\HFD(t)$ is defined in the way that it contains
only degrees of freedom of the formal device.
Therefore $\HFD(t)$ can not mediate any interaction with the stroboscopic
wave packets~(\ref{eq:strobo_notations}) and we have another simplification:
\begin{equation}
\HFD(t) |o;t\rangle = 0 \ \ \textrm{for each} \ o
\ .
\label{eq:HFD_o_zero}
\end{equation}
Hence five of the terms of the original SchE (when written in the considered
mixed basis set) are eliminated.
We now project the resulting equation on the particular basis
vectors, first on wavepackets $|o; t\rangle$ and then on localized atomic
orbitals $|u\rangle$.
By definition, the coupling Hamiltonian $\HFDLC$ fulfils
\begin{equation}
\langle o;t|\HFDLC(t)|o'; t\rangle = 0
\label{eq:o_HFDLC_o_zero}
\end{equation}
and
\begin{equation}
\langle u|\HFDLC(t)|u'\rangle = 0
\label{eq:u_HFDLC_u_zero}
\ .
\end{equation}
(The stroboscopic wave packets $|o;t\rangle$ involve only the degrees
of freedom from the exterior of the formal device
while the basis vectors $|u\rangle$ cover only the formal device.)
Next we examine matrix elements $\langle o;t|\HBias(t)|o'; t\rangle$.
Using definition~(\ref{eq:HBias}), the orthogonality of Hilbert subspaces
of the formal device and of its exterior, and finally employing also the
completeness of the total vector space we find that
\begin{equation}
\langle o;t|\HBias(t)|o'; t\rangle
=
e U_L(t) \delta_{o,o'}
\ .
\label{eq:o_HBias_o}
\end{equation}
In addition it is easily found that [see eq.~(\ref{eq:HBias})]
\begin{equation}
\langle u|\HBias(t)|o'; t\rangle = 0
\ .
\label{eq:u_HBias_o_zero}
\end{equation}
Known matrix elements~(\ref{eq:o_HFDLC_o_zero}), (\ref{eq:o_HBias_o}), (\ref{eq:u_HBias_o_zero})
and (\ref{eq:u_HFDLC_u_zero})
thus allow us to arrive at equations of motion in the form
\begin{equation}
i \hbar \dot{\mathcal{A}}_o(t)
=
e U_L(t) \mathcal{A}_{o}(t)
+
\sum_{u'} A_{u'}(t) \langle o;t| \HFDLC(t) |u'\rangle
\label{eq:for_Ap_prelim2}
\end{equation}
and
\begin{equation}
i \hbar \dot{A}_u(t)
=
\sum_{o'} \mathcal{A}_{o'}(t) \langle u| \HFDLC(t) |o'; t\rangle
+
\sum_{u'} A_{u'}(t) \langle u| \HFD(t) |u'\rangle
\ .
\label{eq:for_Au_prelim2}
\end{equation}
Equations~(\ref{eq:for_Ap_prelim2}) and (\ref{eq:for_Au_prelim2}) are the main results of the Sec.~\ref{sec:method}.
Whereas the number of equations for the amplitudes $A_{u}(t)$ in the finite formal device is finite, 
the number of equations for the amplitudes $\mathcal{A}_{o}(t)$ in the leads is in principle infinite.
However, due to the short range of the matrix elements $\langle o;t| \HFDLC(t) |u'\rangle$, 
only finite and small number of these needs to be solved numerically. This is the unique and efficient 
way the time-dependent semi-infinite leads are treated in our method.

The system under study contains many non-interacting electrons
(satisfying the Pauli principle).
Therefore each probability amplitude will be labeled also by the corresponding electron index:
$\mathcal{A}_{o}^\mathrm{el}$ and $A_u^\mathrm{el}$,
with $\mathrm{el} \in \{1, \dots, \Nel\}$,
where $\Nel$ is the number of explicitly considered electrons in the system.
This number fluctuates in the course of the calculation as the system is an open one.
To choose the initial state we need to specify the amplitudes for all initially considered electrons in the system.
We start from a partitioned state with each of the electrons being initially either in the leads or in the formal
device.

In the leads, the electrons occupy all the stroboscopic states from the few lowest 
energy bands; the energy of the upper limit of the highest occupied band sets the Fermi energy in a given lead. 
Specifically, in our tight-binding implementation of the leads, we use $\Nb = 2$, stroboscopic
wavepackets belonging to band $n=1$ are initially fully occupied, and $n=2$ wavepackets unoccupied.
This occupation corresponds to the filled lower half of the whole TB range of the energies which
is $[\epsilon+2\tB, \epsilon-2\tB]$.\footnote{This is the most
efficient choice. In several test examples we split the total TB energy range into $\Nb = 3$ bands
of unequal widths such that the bottom two bands covered the whole conduction band,
$\eigE \in [\epsilon+2\tB, \, \epsilon]$.}
In the formal device we initially occupy the lower half of the eigenstates of the isolated formal device, which
is then transformed into particular values of the amplitudes $A_u^\mathrm{el}$ for the electrons there.

We solve the set of the differential equations~(\ref{eq:for_Ap_prelim2}) and (\ref{eq:for_Au_prelim2})
numerically using the modified-midpoint method~\cite{numrec}.
Initially the system evolves in time without any bias or gate potential in order
to reach a (quasi)stationary equilibrium states of the coupled system.

During the course of the simulation the unitarily propagating stroboscopic
basis states (wavepackets) are periodically reindexed with the period of
$\tau_n$ given by definition~(\ref{eq:tau_n}).
Along with this, new explicit electrons are inserted into the system and some electrons may be removed
(those which fully or almost fully escaped the explicitly considered Hilbert subspace).
These procedures are done in the same way as in the original SWPA~\cite{our_EPJB} hence we omit their description here.

Having calculated time-dependent probability amplitudes
$\mathcal{A}_{o}^\mathrm{el}$ and $A_u^\mathrm{el}$, we are able to compute
quantities dependent on time and position like the electron density
and the local electron current.
Formula for electron current is of a standard kind
(Ref.~\onlinecite{NEGF}, p.~162 therein)
and have been provided also in our work~\cite{our_EPJB}
[eqs.~(16) and (18) therein].
%
%
\section{\label{sec:graphene}Time-dependent currents in graphene nano-ribbon junctions}
%
%
\subsection{\label{ssec:GNR_overview} Motivation and former work}
%
%
Graphene nano-ribbons (GNRs) are promising building material for a range of future 
nanoelectronic, spintronic and photonic~\cite{Novoselov_Geim,Peres10} devices.
In the following we use the generalized stroboscopic approach to study
coherent time-dependent electronic transport in GNRs.
While first-principles calculation of the quasi-steady-state currents in these systems
is now possible (e.g. in Ref.~\onlinecite{Areshkin10} the NEGF-DFT 
method was used to compute charge transfer, redistribution
and conductance in graphene nano-ribbons up to 7000 atoms), the simulations
of time-dependent transport are rare.
An approach for self-consistent ac quantum transport in the presence of time-dependent 
potentials at nontransport terminals has been applied to a carbon nanotube transistor~\cite{Kienle10}.
The latter method allows to study high-frequency effects also in low-dimensional
nanoscale structures including graphene devices.
The universal and/or non-universal character of conductance and conductivity in GNRs has been 
studied by several groups. 
Katsnelson used the the Dirac-Weyl model of electrons at the neutrality point in graphene samples of a finite size
in one direction and periodic (nanotube-like) boundary conditions in the transverse direction~\cite{Katsnelson06}.
He found the minimum dc conductivity $e^2/(\pi h)$ per valley per spin.
This quantity has been found also by Ludwig \textit{et al.} in their study of the integer quantum Hall effect~\cite{Ludwig1994}.
Tworzyd{\l}o \textit{et al.}~\cite{Beenakker06} pointed attention to the fact that at the neutrality point, 
for $W \gtrsim 4 L$ the dc conductivity attains the value of $\sigma_\mathrm{dc} = (4/\pi) \, e^2/h$ 
(in agreement with Ref.~\onlinecite{Katsnelson06})
and that for smaller aspect ratios it is dependent on this ratio.

On the other hand, in a sequence of papers Lewkowicz,
Rosenstein \textit{et al.}~\cite{Rosenstein09,Rosenstein10R,Rosenstein11}
addressed temporal evolution of the current density in the extended GNRs.
In the linear-response approximation they found~\cite{Rosenstein09} that long-time currents at zero temperature
led to a conductivity value $\sigma_\mathrm{ac} = (\pi/2) \, e^2/h$.
In addition, they have found rapid oscillations in current density with the period $\pi\hbar/\gamma$, $\gamma$ being
the interatomic hopping constant in their notation.
Further analysis of dynamical processes relevant at different time scales for electronic transport in mesoscopic graphene
samples was given in Ref.~\onlinecite{Rosenstein11}
and the dc value $\sigma_\mathrm{dc} = (4/\pi) \,e^2/h$ valid for $W \gg L$ was confirmed.

The time-dependent ballistic transport in metallic GNRs as a response to switching of the bias has been studied
by Perfetto \textit{el al.}~\cite{Perfetto10} for a tight-binding model of a finite GNR attached to identical
semi-infinite GNRs.
According to their findings for large enough widths ($W \gtrsim L \gtrsim 20\,$nm), the time-dependent currents
display two plateaus.
The first plateau occurs at short time scales and corresponds to the $\sigma_\mathrm{ac} = (\pi/2) \,e^2/h$ value of
the conductivity which is the same for both open and closed (a nanotube) boundary conditions, and independent 
of the ratio $W/L$.
At longer times, the current reaches a quasi-stationary value (the second plateau) corresponding to the 
dc conductivity, which depends on the kind of the boundary conditions.
For AGNRs with $W < L$, the authors found long-time quasi-stationary currents with magnitudes corresponding
to the single quantum of conductance, $G_0 = 2 e^2/h$, independently of the ratio $W/L$.

For the purposes of easier comparison, particularly with Ref.~\onlinecite{Perfetto10}, we present results for 
abrupt switching of the bias and study the \emph{armchair} graphene nano-ribbons (AGNR) only, even though the use of 
other time-dependent potentials or other graphene nano-flakes is easily done within our program.
On the other hand, we extend the above studies in several directions:
(i)~apart from the full coupling of the electrodes to the GNRs that has been subject of previous studies we also
examine the reduced-coupling case and discuss its consequences on the temporal character of the current response;
(ii)~we revisit the universality of the long-time quasi steady states for different aspect ratios $W/L$;
(iii)~we explore the difference in the time-dependent current when measured in the electrode or in the graphene
nano-ribbon.
%
%
\subsection{\label{ssec:GNR_model}Model of the nano-ribbon and the electrodes}
%
%
To study the temporal dynamics of the current through graphene nano-ribbon we employ a simple tight-binding model 
with a single orbital per atom~\cite{Wallace}. It is now well established that this model is qualitatively and to some
extent also quantitatively correct~\cite{Louie_PRL06}. Our tests in which we included interatomic hoppings
up to the 3rd nearest neighbors showed only marginal modifications to computed time-dependent currents.

\begin{figure}[!thb]
\centerline{\includegraphics[width=0.49\textwidth]{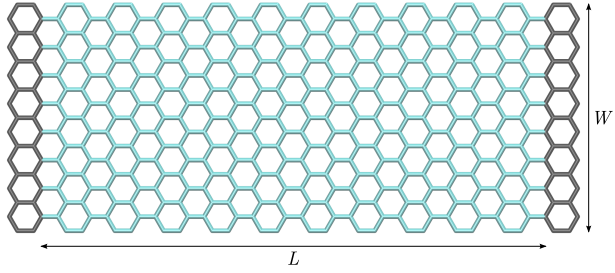}}
\caption{Graphene nano-ribbon with longer edges of the armchair kind.
Its total number of atoms is 408.
Its total length is $L_\mathrm{tot} = 35\,a \approx 49.70\,${\AA} and its width is
$W = 8\sqrt{3}\, a \equiv 8\,b \approx 19.68\,${\AA}
($a$ being the nearest-neighbor distance in graphene.
Its numerical value is in fact irrelevant in our model).
In the notation of Refs.~\onlinecite{Nakada96,Louie_PRL06}, the width of this AGNR corresponds to $N_a = 17$ dimers.
The 68 ($2 \times 34$) black-colored atoms represent the sites where the two electrodes are attached (see text).
In the notation of our method it means $\NL = 68$ and the physical device consists of $M_0 = 408$ atoms.
According to our model, the bias voltage profile inside the GNR has the uniform slope along the length
$L = L_\mathrm{tot}-4a$.}
\label{fig:GNR_img_basic}
\end{figure}

Specifically, the considered structures are armchair graphene nano-ribbons (such as shown
in Fig.~\ref{fig:GNR_img_basic}).
%
%
We consider two electrodes that are attached to the finite AGNR: the source electrode contacted to the left 
side of the GNR and the drain electrode attached to the right side.
Our model of an electrode is a bundle of the identical semi-infinite mono-atomic chains
described as leads in the SWPA (Sec.~\ref{sec:method}), interacting via the nearest-neighbor (NN) TB hopping 
parameter $\tB$.
The black color in Fig.~\ref{fig:GNR_img_basic} marks those atoms of the GNR to which the mono-atomic chains
(composing particular electrode) are attached.
Formally, the entire system is a multi-terminal structure defined in Sec.~\ref{sec:method}
(see Fig.~\ref{fig:theory_junction}).
This simple model of the electrodes covers the physical aspects of realistic contacts only partially, but it is
sufficient for the present qualitative model of the dynamical transport.

The NN approximation is used also for the graphene nano-ribbon.
To keep the model simple, we use the same value of the hopping parameter $\tB$ for both - the chains and the GNRs.
The exception from the single-hopping value model is the contact between the chains and the GNR described by the
parameter $\tGE$ which is generally different from $\tB$.

Our simulation protocol consists of three subsequent steps:
(i)~An initial many-electron non-interacting state is defined.
Its description is provided in subsection~\ref{ssec:SchE} below eq.~(\ref{eq:for_Au_prelim2}).
(ii)~The entire system is allowed to evolve according to the SchE at zero bias for a sufficiently long interval
of time ($300\,|\tB|/\hbar$ in the present work).
We call this stage \emph{equilibration}.
During the equilibration the electronic structure is adapted to the given zero-bias potential generated
by the electrodes and by the GNR nanojunction.
(iii)~At time $\tsw = 300\,|\tB|/\hbar$ we abruptly turn on the bias voltage $U$, which is then kept constant.
The effect of the electric bias is modeled by the lift of the on-site energies in the whole source electrode (including
the finite parts of the electrodes belonging to the formal device) by the amount of $e U$.
The on-site energies of the drain electrode ($\epsilon$) are unchanged.
In addition, the bias has its effect on the on-site energies of the physical device (here the GNR).
In an \textit{ab-initio} model we would have to compute the effect using at least a self-consistent field method.
Because in the present work we use the independent-electron approximation, we have to prescribe a model of the
bias-induced variation of the on-site energies.
We assume the linear variation of the bias profile within the GNR.
However, each atom which is directly contacted to one of the mono-atomic wires (the black-colored atoms on
Fig.~\ref{fig:GNR_img_basic}), is kept at the same orbital energy as the atoms in the particular wire (chain).
Although we consider relatively large bias voltages (typically $e U = 0.5\,|\tB|$) they are still significantly lower
than the bandwidth in the leads which is $4\,|\tB|$.

We compute the local electric current at a range of bonds of the mono-atomic wires using formulas provided
in Ref.~\onlinecite{our_EPJB} (see also Ref.~\onlinecite{NEGF}, p.~162 therein).
Because a single electrode is composed of a finite number of the wires, we sum up the currents over the wires
of particular electrode and obtain the total current through the electrode.
We also compute the current inside the GNRs.
%
%
\subsection{\label{ssec:full_coup} Full coupling, single dominant channel}
\begin{figure}[!thb]
\centerline{\includegraphics[width=0.49\textwidth]{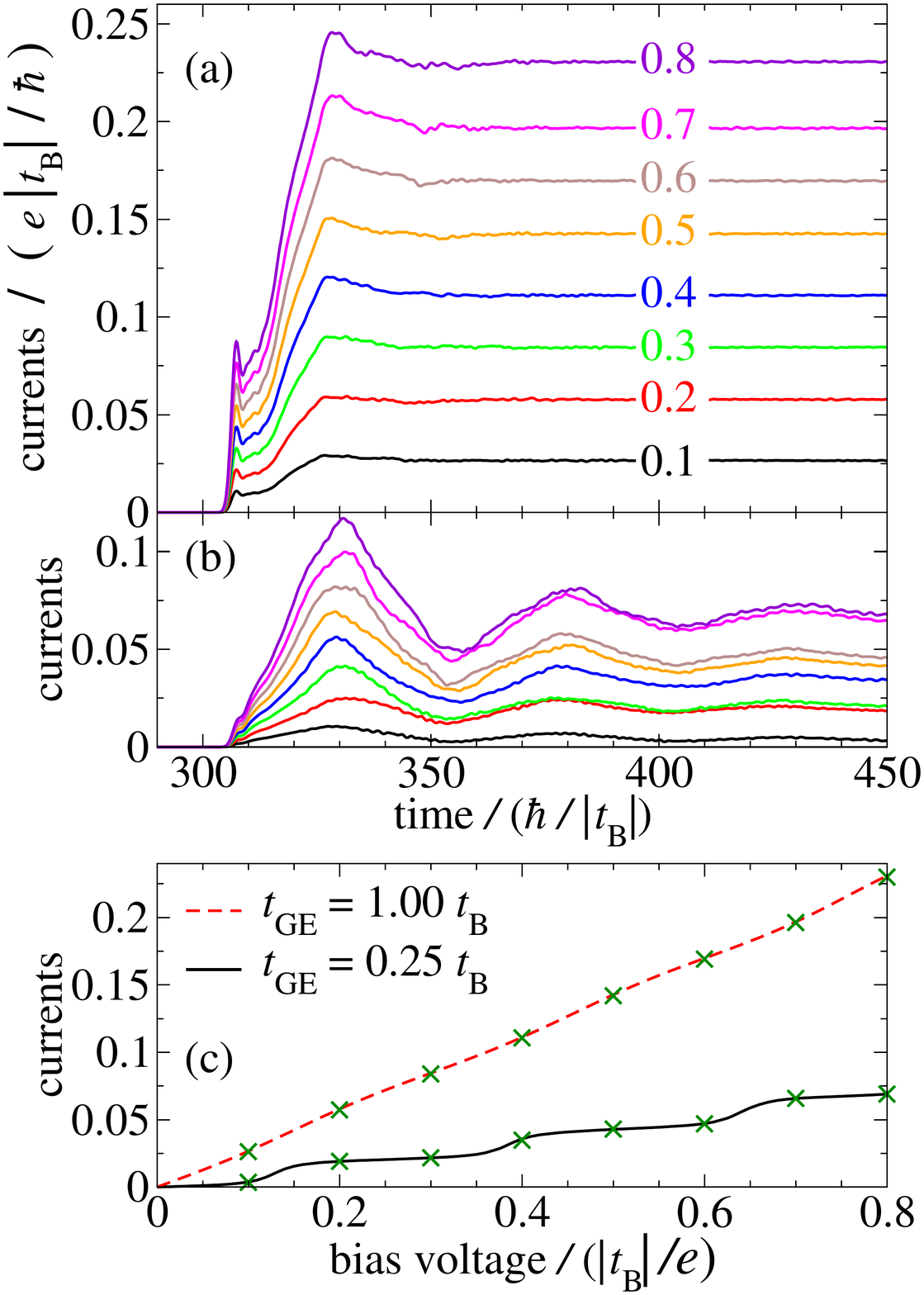}}
\caption{\textbf{(a)}~Time-dependent electron currents through the nano-junctions formed by the AGNR shown
on Fig.~\ref{fig:GNR_img_basic}.
The coupling parameter between the GNR and the electrodes is $\tGE = \tB$.
Particular plots correspond to bias voltages $U = 0.1$, $\dots$, $0.8$ in units of $|\tB|/e$.
The voltages were turned on abruptly at time $\tsw = 300\,\hbar/|\tB|$.
\textbf{(b)}~Similar as in the (a) case but with $\tGE = 0.25\,\tB$.
Vertical ordering of the plots is the same as in graph~(a).
\textbf{(a,b)}~The plotted currents have been calculated in the drain electrode.
The used basis set corresponds to $\mmax=100$ (the stroboscopic part), $M_0 = 408$ (the number of localized orbitals
in the GNR) and $M_1 = M_2 = \dots = M_{68} = 40$ (the number of the localized orbitals in the finite piece
of each electrode).
\textbf{(c)}~Line plots show stationary current-voltage characteristics calculated with an exact numerical
eigenstate analysis based on Green's functions~\cite{Ryndyk} (see also main text).
The isolated x symbols show values of the long-time quasi-stationary currents from the dynamical simulations 
similar to those shown in (a) and (b) (with $\mmax=50$ and $M_1 = M_2 = \dots = M_{68} = 20$)~\cite{reduced_basis}.}
\label{fig:GNR_Itdep_Biases}
\end{figure}

\begin{figure}[!thb]
\centerline{\includegraphics[width=0.49\textwidth]{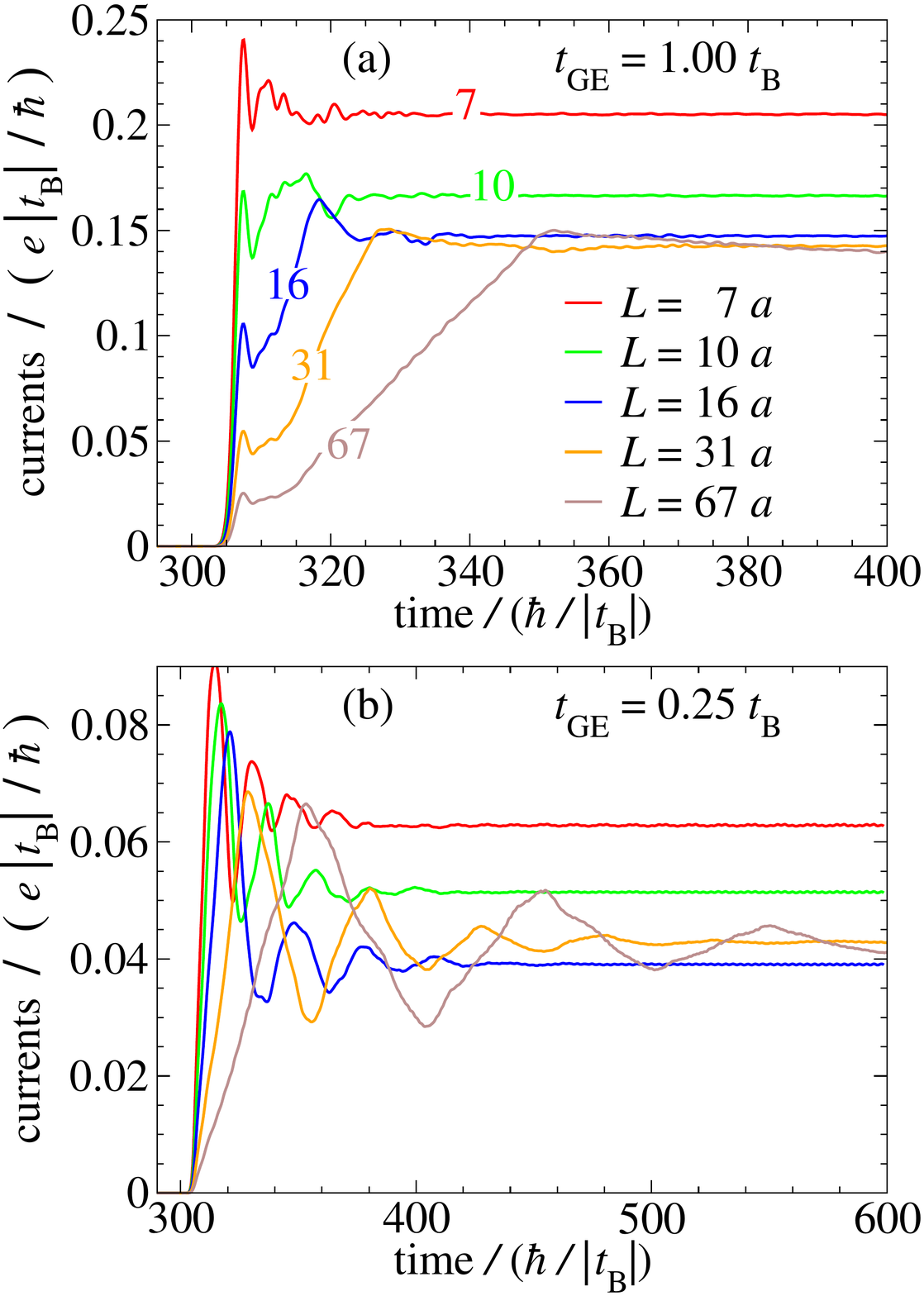}}
\caption{Time-dependent electric currents through the nano-junctions formed by the AGNRs of five different lengths
$L$ and of the same width $W = 8\,b$ corresponding to $N_a = 17$ (same width as shown on Fig.~\ref{fig:GNR_img_basic}).
Bias voltage $U = 0.5\,|\tB|/e$ was turned on abruptly at time $\tsw = 300\,\hbar/|\tB|$.
As in Fig.~\ref{fig:GNR_Itdep_Biases}, the plotted currents have been calculated in the drain electrode at the distance $10.5\,a$ from the AGNR.
Graphs (a) and (b) differ by the GNR-electrodes coupling strength $\tGE$, as indicated in the respective graphs.
The vertical ordering of the plots in (b) at initial times is the same as in (a).
The results were obtained using $\mmax=100$ and $M_1 = M_2 = \dots = M_{68} = 40$.
Running average over the interval of $\tau = \pi\hbar/|\tB|$ was used in (b)
to remove small unphysical oscillations caused by the finite $\mmax$ value~\cite{runav}.}
\label{fig:GNR_Itdep_Lengths}
\end{figure}

%
%
In the present subsection we discuss results for the GNRs fully coupled to the electrodes, i.e. we use $\tGE = \tB$.
This corresponds to the case studied in previous works even though it is less realistic in view of a typical 
realizations of a contact between GNR and electrodes via a tunneling barrier~\cite{Wang2008}.
Within this subsection,
the geometrical arrangement of the GNRs have in general long and narrow shape, i.e. $W \ll L$,
which will result in a transport through a single dominant channel in the long-time limit. 
We use several values of the bias voltage $U$ which allows us to compare $I(t)$'s for different $U$'s.
Dynamical currents from the simulations are plotted in Fig.~\ref{fig:GNR_Itdep_Biases}a.
%
%
The plotted values of the currents were evaluated in the drain electrode, in between sites 10 and 11 counted
from the GNR.
Because we will make references to the works~\onlinecite{Rosenstein09,Rosenstein10R,Perfetto10,Rosenstein11} we remark
that those authors computed their dynamical currents in the middle of the GNR.
In addition, our model of the electrodes is different
and our results are obtained in a higher bias voltage regime.
Therefore our results are not directly comparable to the results of those authors.
However, in section~\ref{ssec:Imid} we will attempt for a closer comparison by presenting currents evaluated
in the middle of the GNR.
In the following paragraphs we discuss particular features of the $I(t)$ functions of Fig.~\ref{fig:GNR_Itdep_Biases}a.
%
\paragraph{The initial increase of the current until the first peak.}
%
The initial extremely fast increase occurs on the timescale of $\hbar/|\tB|$ which is the natural timescale in pure
graphene~\cite{Rosenstein09,Perfetto10,Rosenstein11}.
Due to dispersion,
the duration of this process varies with the chosen position where the dynamical
current is computed.
Immediately next to the GNR the duration is about $1.5\,\hbar/|\tB|$.
Fig.~\ref{fig:GNR_Itdep_Biases} shows time-dependent currents computed between sites 10 and 11.
However, most of the temporal features are independent of chosen observation point at least in the range up to $200\,a$
from the GNR.
The initial extremely fast process ends at a peak current value $I_\mathrm{peak}$.
The dependence $I_\mathrm{peak}(U)$ extracted from our data is perfectly linear.
%
\paragraph{The minimum after the peak.}
%
The initial increase and the peak in the $I(t)$ curves on Fig.~\ref{fig:GNR_Itdep_Biases}a is followed by the short
dip which would extend to a plateau for smaller bias voltages.
Inspection of the dynamical currents from our simulations shows that this value of the current, denoted as
$I_\mathrm{plat}$, is again linear with respect to the applied bias voltage $U$, with the coefficient of
determination $R_\mathrm{plat}^2 = 0.999916$.
The slope of the linear dependence yields the conductivity
\begin{equation}
\sigma_\mathrm{plat} = \frac{I_\mathrm{plat}}{U} \, \frac{L}{W} \approx 0.86\,\sigma_\mathrm{ac}
\ ,
\end{equation}
where $\sigma_\mathrm{ac}$ is the theoretical value~\cite{Rosenstein09} equal to the graphene ac conductivity:
\begin{equation}
\sigma_\mathrm{ac} = \frac{\pi}{2} \, \frac{e^2}{h} \, .
\label{eq:sigmaac}
\end{equation}
The plateau effect has been discussed also in Ref.~\onlinecite{Perfetto10} for GNRs.
Again, the differences in the studied models, especially different models of the electrodes, do not allow for detailed
direct comparison of the result in Ref.~\onlinecite{Perfetto10} and our present work.
The lower value of $\sigma_\mathrm{plat}$ compared to $\sigma_\mathrm{ac}$ is mainly due to these differences.
We also find that currents computed \emph{closer} to the GNR tend to display more plateau-like transient effect.
The plateau effect is also enhanced for longer GNRs (larger $L$), as can be seen from
Fig.~\ref{fig:GNR_Itdep_Lengths}, where it appears due to smaller bias-induced electric field $U/L$.
%
\paragraph{The massive quasi-linear increase of the current.}
%
This process occurs beyond the linear-response regime~\cite{Rosenstein10R,Rosenstein11}.
Its duration is directly proportional to the GNR length; the global maximum of
the current is reached at the time $L/v_\mathrm{FN}$ after the bias turn-on,
where $v_\mathrm{FN}$ is the Fermi velocity in the nano-ribbon under consideration,
$v_\mathrm{FN} \approx 1.5\,a |\tB|/\hbar$, which is also the Fermi velocity in ideal
graphene~\cite{Wallace,Novoselov_Geim}.
See also subsection~\ref{sssec:lengths} and especially Fig.~\ref{fig:GNR_Itdep_Lengths}a for a clear demonstration.
The massive increase of the current can roughly be understood in quasi-classical terms: the Bloch wavevectors
of electrons within the GNR move in the Brillouin zone under the constant force $e U/L$.
The transport is ballistic (no inelastic scattering) which results in the uniform increase of the current during the
initial traverse time through the GNR.
The global maximum value of the current extracted from our simulations is only marginally dependent on the chosen point
in the electrodes (we tested distances up to $40.5\,a$ from the GNR).
The referred maximum is global only in regime $W \lesssim L$.
For example,
results in Figs.~\ref{fig:GNR_Itdep_Lengths}a and \ref{fig:GNR_Itdep_Widths}a show that the opposite aspect ratios, 
$W/L > 1$, 
provide the global maximum already in the first peak of the $I(t)$ dependence at the time of the order $\hbar/|\tB|$
after the bias voltage turn-on.
%
%
Compared to other results in the literature (Ref.~\onlinecite{Perfetto10}, Fig.~5 therein),
our model exhibits smoother $I(t)$ curves due to our choice to compute the current in the lead, not in the GNR.
%
\paragraph{Partial decrease of the current after the maximum.}
%
The partial relaxation of the maximum current occurs again on the timescale of $L/v_\mathrm{FN}$.
The relaxation is due to the adapted electronic state of the GNR.
While in the steady state only the electron amplitudes with the energies from the bias window
$[\epsilon, \epsilon+e U]$ contribute to the total current, during the transient effect the
remaining portions of the spectrum are also relevant.
%
\paragraph{Long-time quasi-stationary behavior}
%
Shortly after the relaxation period the local current in the electrodes reaches a quasi-stationary limit
(Fig.~\ref{fig:GNR_Itdep_Biases}a).
The long-time value of the current can also be obtained from a stationary analysis such as the one used in the
Landauer-B\"{u}ttiker formalism and the transfer matrix.
(Our model of the electrodes formally corresponds to a multiterminal device.)
We do such an analysis and obtain stationary current-voltage characteristics which we compare to the
quasistationary currents from the dynamical simulations.
To this end, we perform a numerical eigenstate analysis similarly as in Ref.~\onlinecite{our_EPJB}, now with the
eigenstates calculated with the Green's function approach applied to our TB model.
(See Ref.~\onlinecite{Ryndyk} for the presentation of the theory in case of systems with two leads.)
The stationary currents $I_\mathrm{stat}$ for the fully coupled AGNR ($\tGE = \tB$) are shown as the red dashed
line in Fig.~\ref{fig:GNR_Itdep_Biases}c.
The green x symbols at the plot are the independently obtained quasi-stationary long-time values extracted from the
dynamical currents of Fig.~\ref{fig:GNR_Itdep_Biases}a.
The agreement is excellent, despite the limited basis set size used in the dynamical simulations.
The impact of the finite basis set is mainly to introduce artificial (unphysical) oscillations to the computed
currents~\cite{our_EPJB} but the currents oscillate around correct values.
If the unphysical oscillations are too large we smooth them by application of running averages~\cite{runav}.
We do it only in a few cases of the reduced coupling $\tGE$ (but not in Fig.~\ref{fig:GNR_Itdep_Biases})
which will be discussed below and indicated in captions to affected figures.
The $I_{\mathrm{large}\ t}(U)$ dependence (now commented for the full-coupling case, Fig.~\ref{fig:GNR_Itdep_Biases}a)
is roughly linear in the studied range of biases:
$I_{\mathrm{large}\ t} = 0.89\,(2 e^2/h) \, U$
obtained by the linear regression over symmetrized data (which include the negative biases).
The deviation from the linearity is most pronounced at the lowest biases where we would have a transmission
coefficient $0.779$ instead of $0.89$.
With a high accuracy we have $I_\mathrm{stat}(U) = I_{\mathrm{large}\ t}(U)$ hence these numbers are valid
for both -- the stationary analysis and the quasi-stationary currents from the simulations.
Relation between $U$ and $I_\mathrm{stat}$ will be further discussed below in connection with the AGNR width.
%
%
\subsection{\label{sssec:lengths}Importance of the aspect ratio for the full coupling}
%
%
In this subsection we consider AGNRs of several different lengths $L$ and widths $W$, with variable aspect ration $W/L$.
%
\begin{figure}[!thb]
\centerline{\includegraphics[width=0.49\textwidth]{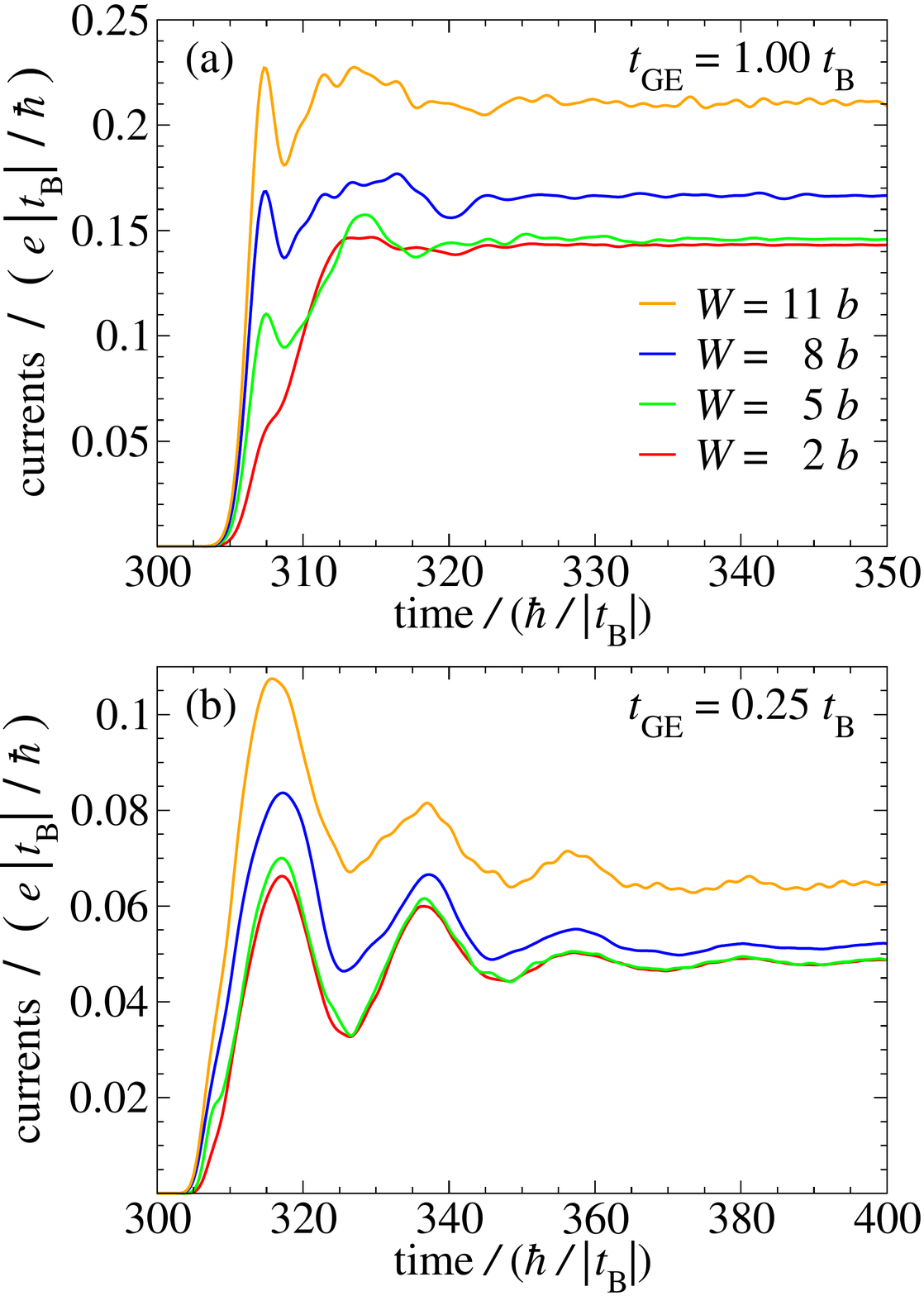}}
\caption{Time-dependent electric currents through the nano-junctions formed by the AGNRs of four different widths
$W$ and of the same length $L = 10\,a$
(see also Fig.~\ref{fig:GNR_img_basic} and caption to Fig.~\ref{fig:GNR_Itdep_Lengths}).
Vertical ordering of the plots in graph (a) is the same as the ordering of the legends.
Vertical ordering of the plots in (b) is the same as in (a) at initial times.
Time dependencies for widths $2$, $5$, and $8\,a$ were obtained using $\mmax=100$ and $M_1 = M_2 = \dots = M_{68} = 40$.
The widest AGNR employed half-sized basis set parameters $\mmax$, $M_1$, \dots, $M_{68}$ compared to those above.
The running averages were applied in graph (b) to plots $W = 8$ and $11\,b$ to remove the non-physical 
oscillations~\cite{runav}.}
\label{fig:GNR_Itdep_Widths}
\end{figure}
%
There are three remarkable features of quantum transport through an \emph{ideal} graphene nano-ribbon 
regarding its dependence on these geometric parameters.
First, the ballistic dc conductance $G$ at the neutrality point is in general a non-linear function of the
aspect ratio $W/L$ only~\cite{Beenakker06}, which is consequence of the absence of any energy scale of the system
at the neutrality point~\cite{Rossi11}.
The dependence on $W/L$ only holds for any~\cite{symmetry} small or large $W$ and $L$.
Second, for $W \gg L$ the dc \emph{conductivity} acquires a constant (universal) value
$\sigma_\mathrm{dc} = \sigma_\mathrm{min} = (4/\pi)\,e^2/h$,
again independent on both $W$ and $L$. 
For this reason many authors refer to the conductivity $\sigma \equiv G L/W$ rather than conductance $G$ when 
discussing coherent transport in graphene and refer to this as the Ohmic-like regime.
Third, for $W \lesssim L$ in contrast with the previous case it is the \emph{conductance} that the takes 
a constant, aspect-ratio-independent value, of the order of $G_0 = 2 e^2/h$. 
In the words of mesoscopic transport, only one channel is active through such AGNRs. 
The universality of the conductance value $G_0$ extends through the whole range of AGNRs having $W \lesssim L$ meaning
that in ideal long GNRs the transport has the typical quantum ballistic character as opposed to the pseudo-Ohmic
behavior of the wide GNRs.

However for systems in which we contact the GNR through a tunneling barrier, i.e.
when $\tGE$ is different from $\tB$, the above scheme is not necessarily valid.
In Appendix~\ref{app:WLrat} we provide plots of the conductance vs $W/L$ ratio for the systems under study
obtained from our stationary calculations.

Our numerical calculations confirm the above described scenario. 
All of the considered GNRs are of the same symmetry~\cite{symmetry} as the GNR plotted on Fig.~\ref{fig:GNR_img_basic}.
Also, all of the considered AGNRs are metallic.\footnote{AGNR is metallic if it has the number of the dimer lines
$N_a = 3p-1$ where $p$ is a whole number~\cite{Nakada96}.}
Each such GNR has the electrodes contacted in the same way as shown on the figure, i.e. the contacted atoms form the
strips across the whole width of the GNR.
We use the simulation protocol described above, choosing the bias voltage $U = 0.5\,|\tB|/e$.
Dynamical currents from the simulations are plotted in Figs.~\ref{fig:GNR_Itdep_Lengths}
and~\ref{fig:GNR_Itdep_Widths}.

Electric currents plotted on Figs.~\ref{fig:GNR_Itdep_Lengths}a and \ref{fig:GNR_Itdep_Widths}a correspond to a rather
high bias $U = 0.5\,|\tB|/e$.
Hence their long-time quasi-stationary values do not exactly depend on $W/L$ only.
Nevertheless, the quasi-stationary currents still exhibit similar features, namely that for $W \lesssim L$ they
are independent on $W$ and $L$ and attain the value $I_{\mathrm{large}\ t}$ which is close to  $(2 e^2/h) \, U$.
%
%
\subsection{\label{ssec:redu_coup}Reduced coupling}
%
%
In this subsection we discuss results for the reduced coupling, $\tGE = 0.25\,\tB$, between the AGNRs
and the electrodes. 
All other settings for the simulations are the same, with few unimportant exceptions like different simulations times.
The reduced $\tGE$ results in tunneling barriers for electrons, causing partial isolation of the GNR
from the electrodes.
The first set of the results are the time-dependent currents shown in Fig.~\ref{fig:GNR_Itdep_Biases}b.
Compared to the fully coupled GNRs, the systems with the tunneling barrier exhibit generally lower currents which is
not surprising.
The most distinctive feature of the $I(t)$s are the damped, triangle-shaped oscillations. This shape, similarly to the
massive quasi-linear increase of the current in the fully coupled system, can be explained by the quasi-classical 
argument with the shifting of the occupied Bloch states in GNR.
The period of the oscillations on Fig.~\ref{fig:GNR_Itdep_Biases}b is independent of the applied bias voltage.
Inspection of the results for the GNRs of different lengths shows that the period is twice the
traverse time of the electrons through the length of the GNR: $T \approx 2 L/v_\mathrm{FN}$
(see Fig.~\ref{fig:GNR_Itdep_Lengths}b).
This is a clear evidence that the damped oscillations originate from the partial reflections of the electron
amplitudes at the ends of the AGNR.
Another interpretation is that the electrons traverse through the resonant eigenstates of the AGNR.
Although the period $2 L/v_\mathrm{FN}$ of the oscillations may look obvious, generally there can be several
kinds of oscillations in time-dependent currents through partially isolated nanojunctions.
For example, damped oscillations can obey the equality $\hbar \omega = e U/2$, which was analytically derived
in Ref.~\onlinecite{JianWang10} and found also in our recent work~\cite{our_EPJB} for quantum rings.
See also appendix~\ref{ssec:circulating}.

The absolute maxima of the $I(t)$ curves in Fig.~\ref{fig:GNR_Itdep_Biases}b are achieved at the first peak,
appearing at the time of about $L/v_\mathrm{FN}$ after the bias voltage had been switched on.
The maximum current scales linearly with the applied voltage.
This is in contrast with the long-time quasi-stationary (or truly stationary) currents which depend on $U$ non-linearly,
as can be seen also from Fig.~\ref{fig:GNR_Itdep_Biases}c, the solid-line black plot.
The appearance of the (rounded) steps in the $I_\mathrm{stat}(U)$ plot reflects the presence of relatively sharp peaks
in the transmittance $T(\mathcal{E})$ (not shown in graphs).

Importantly, also in the systems with the reduced coupling $\tGE$, the long-time values of the currents for long narrow
AGNRs have typical quantum ballistic dependence on the GNR's dimensions, i.e. the currents at large times do not
depend on neither $W$ nor $L$ (if $W < L$).
This can be seen from Figs.~\ref{fig:GNR_Itdep_Lengths}b and.~\ref{fig:GNR_Itdep_Widths}b.
In Appendix~\ref{app:WLrat} we provide more detailed stationary results.
To the best of our knowledge, GNRs with the reduced coupling to electrodes and resulting oscillations have not been
considered in existing literature.
%
%
\subsection{\label{ssec:Imid}Currents evaluated in the middle of the GNR}
\begin{figure}[t]
\centerline{\includegraphics[width=0.49\textwidth]{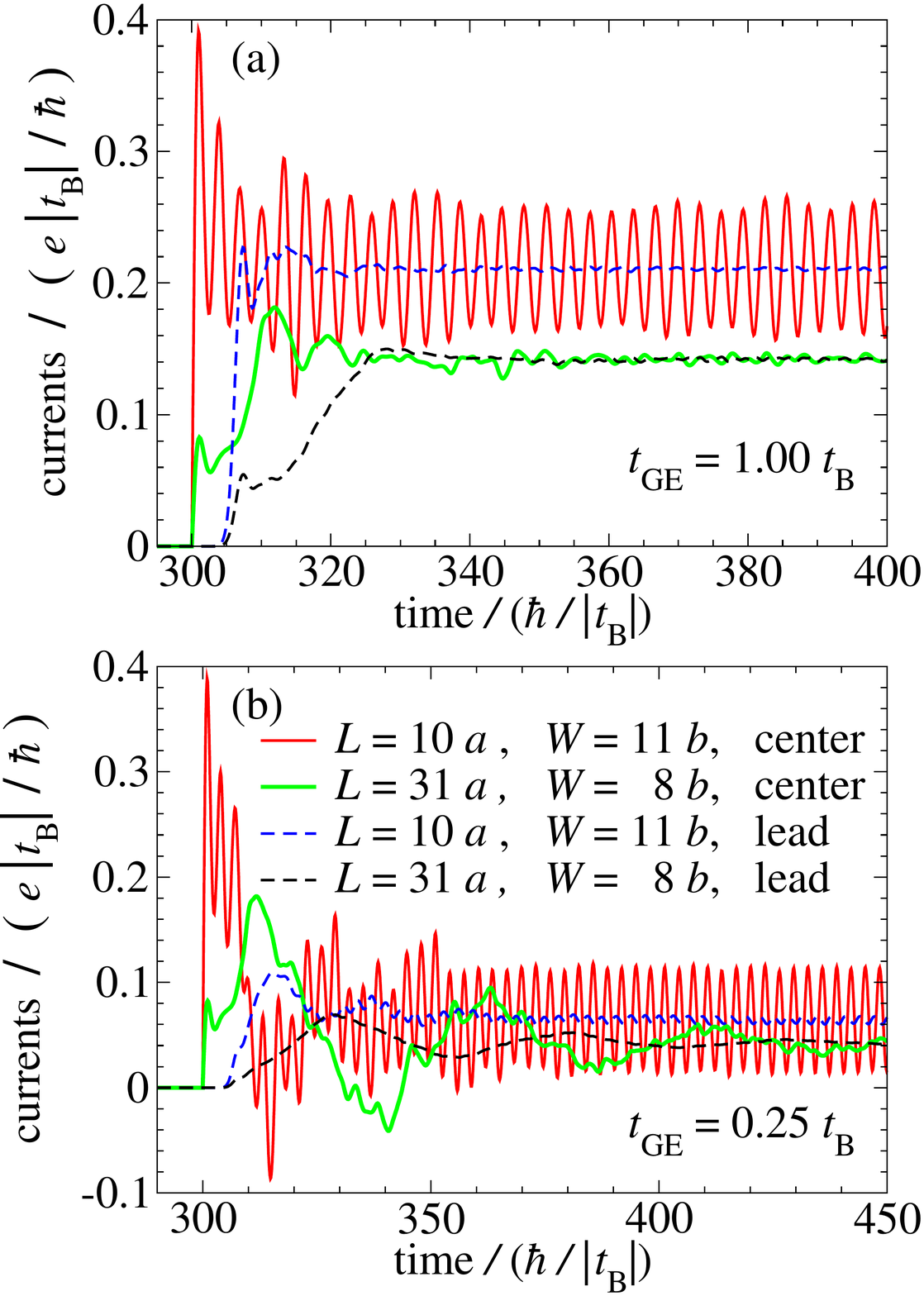}}
\caption{Comparison of time-dependent currents evaluated in the centers of the AGNRs and in the drain electrode (at the
distance of $10.5\,a$ from the GNRs).
(a)~The structures with the full GNR-electrodes coupling, $\tGE = \tB$.
(b)~The structures with the reduced GNR-electrodes coupling, $\tGE = 0.25\,\tB$.
All currents were obtained for the abruptly turned-on bias voltage $U = 0.5\,|\tB|/e$ 
under the same conditions as used for the results on Figs.~\ref{fig:GNR_Itdep_Lengths} and \ref{fig:GNR_Itdep_Widths}.
The delay between the first increase of the currents evaluated in the centers and in the drain electrode is 
due to finite propagation of the density disturbance in the electrode given by the Fermi velocity.
No additional smoothening was applied to the data.
We stress that the shown rapid oscillations (the red plots) are not an artifact of the finite basis set.}
\label{fig:GNR_Itdep_Mid}
\end{figure}

%
%
All currents presented above have been calculated in the drain electrode at the distance of $10.5\,a$ from the AGNR.
Other works on time-dependent transport in GNR nanojunctions consider currents in the middle of the GNR,
see for instance~\cite{Perfetto10,Rosenstein11}.
In Fig.~\ref{fig:GNR_Itdep_Mid} we compare the time-dependent currents calculated in the middle of AGNRs
and in the lead for two types of AGNRs:
(i)~long-narrow structures ($L > W$, here $L = 31\,a$, $W = 8\,b$) and
(ii)~short-wide structures ($L < W$, here $L = 10\,a$, $W = 11\,b$).
%
%
The difference given by the two qualitatively different observation points -- the center and the electrode -- is
striking in several of the presented cases and is typical also for other AGNRs.
Namely, the large rapid oscillations with the period of $\pi\,\hbar/|\tB|$ (red solid plots
on Fig.~\ref{fig:GNR_Itdep_Mid}) are found only inside the AGNR, not in the leads.
Similar oscillations have been found also in works~\cite{Rosenstein09,Perfetto10}.
Inspection shows that the oscillations occur only if the local current is evaluated in the wider cross section
of the AGNR.
For example, the structure shown on Fig.~\ref{fig:GNR_img_basic} has its (wider) width $W = 8\,b$, but locally the
width alternates between two values: $8\,b$ and $7\,b$.

In addition to the rapid oscillations, the currents recorded in the middle of the AGNRs can exhibit additional
aperiodic oscillations as can be seen by comparing the currents for the $L = 31\,a$ cases on
Fig.~\ref{fig:GNR_Itdep_Mid} (solid green lines vs dashed black lines).
In the case of the full coupling (Fig.~\ref{fig:GNR_Itdep_Mid}a) of the longer AGNR (solid green plot) we can make
almost direct quantitative comparison of our current evaluated in the center with the work
of Perfetto \textit{et al.}~\cite{Perfetto10}.
Specifically, Fig.~5 in their work shows time-dependent currents also for the AGNR of the width
$W_2 = 2.1\,\mathrm{nm}$ which is similar to our AGNR of the width $W = 8\,b \approx 1.97\,\mathrm{nm}$.
The lengths of the two AGNRs are also similar.
Comparison of the results shows clear quantitative agreement ($W_2 = 2.1\,\mathrm{nm}$ plot in
Fig.~5 of Ref.~\onlinecite{Perfetto10} vs. the green solid plot in our Fig.~\ref{fig:GNR_Itdep_Mid}a).
In our case the peak value of the current is about $I_\mathrm{peak} = 0.19\,e |\tB|/\hbar$ which translates to
$I_\mathrm{peak}/(U\,\sigma_0) = 2.39$, with $\sigma_0 \equiv e^2/h$.
In Ref.~\onlinecite{Perfetto10} we find the peak value about $2.1$ which compares with our value $2.39$ well.
A precise equality of the $I(t)$ curves can not be expected due the differences in the models of the electrodes
and in addition we consider the AGNR of a slightly different size.

Even if we disregarded the rapid oscillations, we would still have noticeable differences between the transient
currents recorded in the center of the AGNR and in the electrode.
The generally more oscillatory $I(t)$ curves in the center are associated also with internal reflections 
of the electrons inside the nano-ribbons.
If the AGNR is partially isolated from the electrodes due to the reduced hoppings $\tGE$
(Fig.~\ref{fig:GNR_Itdep_Mid}b) the oscillations at the period of $L/v_\mathrm{FN}$ are strongly enhanced.
Hence we conclude that transient currents evaluated within a GNR are generally different from currents
obtained in the electrodes and that only the former ones noticeably reflect internal charge density oscillations in the
ribbon.
%
%
%
\section{\label{sec:conclu}Discussion and conclusions}
%
%
We have presented the generalized stroboscopic wavepacket approach to electronic quantum transport and used
it to study time-dependent currents in graphene nano-ribbon (GNR) junctions.
The generalized method employs a mixed basis set composed of the stroboscopic wavepackets and from the localized
atomic orbitals.
The wavepackets are used to describe wavefunctions in the semi-infinite leads while the localized functions cover the
central part of the system.
In its present formulation the approach uses tight-binding Hamiltonians and the independent-electron model.
The method presents a major improvement over our previous approach~\cite{our_EPJB} which was based on purely
stroboscopic basis functions.
The improvements are four-fold:
(i)~The entire system can include arbitrary number of attached leads or terminals.
(ii)~The method allows to study arbitrarily structured nano-junctions.
(iii)~The accuracy (convergence of results with respect to the basis set size) allows to obtain very good
convergence at affordable basis set sizes.
(iv)~The required computational resources, especially computer time, has
been lowered by several orders of magnitude.
Even the most memory-demanding calculations required no more than 52 GiB of operating memory.
The bulk of calculations have been obtained with a 12-core machine.

We can thus conclude that the stroboscopic wavepacket approach (SWPA) in its generalized form presents a viable
alternative to the non-equilibrium Green's function (NEGF) method for time-dependent quantum transport.
The SWPA can be shown to be in principle equivalent to the NEGF-TDDFT method but brings the advantage of uncomplicated
incorporation of the semi-infinite leads.

In the study of the time-dependent currents through the GNR junctions we have considered armchair GNRs (AGNRs) and the
effect of the coupling strength between the GNR and the electrodes.
As opposed to other theoretical or computational studies in the literature, we determined electric currents mainly in
the electrodes, not only in the GNR itself.
We studied the transport at several different bias voltages, the voltage being abruptly turned on.
The range of voltages was chosen such that a significant portion of the available conduction band width was probed.
The resulting electric fields are rather high compared to typical stationary situations but comparable to 
the fields used in recent studies of ultra-fast electron dynamics~\cite{Tani12}. 
Our study covers both long-narrow AGNRs (dc conductance of which takes a universal value close to $G_0 = 2e^2/h$)
and also short-wide AGNRs [dc conductivity of which takes a universal value close to
$\sigma_\mathrm{dc} = (4/\pi) \, e^2/h$].
The considered sizes extend up to about $10\,\textrm{nm}$ in length and $2.7\,\textrm{nm}$ in width.
Our main results for the AGNRs can be summarized as follows.

(i)~In the cases of the reduced GNR-electrodes coupling we have found damped, triangle-shaped oscillations in the
current which have the period $2L/v_\mathrm{FN}$, $v_\mathrm{FN}$ being the electron propagation velocity in graphene
at the Dirac point.
The maximum peak current obtained at about the time $L/v_\mathrm{FN}$ scales linearly with the applied bias voltage,
including the regime of high bias-induced electric fields;
the studied range of voltages extends up to about $2.3\,\textrm{V}$.

(ii)~In the case of the full GNR-electrodes coupling 
the results for our model of the electrodes are qualitatively 
similar to the results obtained by other authors, e.g. Ref.~\onlinecite{Perfetto10}.
Quantitatively, the currents and dc conductivities have been found to be typically $10-15\%$ lower compared 
to the other models of electrodes.

(iii)~We have compared the time-dependent currents in the electrodes and in the middle of the AGNRs.
The currents inside the GNRs exhibit generally more oscillatory character which is caused by multiple partial
reflections of the electrons inside the GNRs.
In the cases of the reduced coupling between the GNR and the electrodes, the damped oscillations of the current have
significantly larger amplitude compared to the oscillations in the electrodes.
Hence taking the interior of a GNR as an observation point (as it was done in other works) may not be
adequate for discussion with those experiments which consider currents in leads.

(iv)~Although the present work was focused on the time-dependent currents, we paid attention also to the limiting case of
large times and examined how the dc conductance of AGNRs with the reduced coupling to the electrodes depends on the
dimensions $W$ and $L$ of the AGNR.
While it is known that for the full-coupling case $G = G(W/L)$ (and importantly, this relation is non-linear in $W/L$
for $W \lesssim L$), we also discussed the dependence $G(W,L)$ for the reduced-coupling case.
Our calculated results for reduced coupling show that for $W/L>1$ the conductance $G(W,L)$ does not depend 
on the ratio $W/L$ only.
%
%
\section{Acknowledgements}
%
%
This work was supported in parts by
the Slovak Research and Development Agency under the contract No.~APVV-0108-11
and by
the Slovak Grant Agency for Science (VEGA) through grant  No.~1/0372/13.
Computer time was provided by the Slovak Infrastructure
for High-Performance Computing (SIVVP, project code 26230120002).
\appendix
%
%
\section{\label{sec:rings}Currents in quantum ring nanojunctions with two leads}
%
%
Recently we have used the original SWPA for ring nanostructures
with two leads and described them in the TB approximation~\cite{our_EPJB}.
The basis set to expand electron wavefunctions consisted purely from the
stroboscopic wavepackets.
The noticeable and discussed feature of the results in Ref.~\onlinecite{our_EPJB}
was the non-satisfactory convergence of the electron current, 
getting worse for higher biases.
More precisely, the impact of the finite basis set on the time-dependent
currents was two-fold: 
(i)~Computed currents exhibited unphysical oscillations of the period of
$\tau_n$ given by eq.~(\ref{eq:tau_n}).
It was possible to smooth the oscillations in post-processing
by application of proper running averages, thus
the oscillations did not present a substantial difficulty.
(ii)~The much more serious issue was that even the smoothed values of currents
were not converged and their values were typically 5-10\% lower compared to
accurate ones (for given model).
Satisfactory convergence at higher biases would require an enormous number of
the basis functions, completely beyond our computational capabilities.
In this section we demonstrate that our present approach removes most of
the convergence issues and we are able to obtain converged results
at moderate basis set sizes.
The unphysical oscillations generally remain but they are usually lower and
could be sufficiently suppressed using a larger but still achievable basis set
size.
As in Ref.~\onlinecite{our_EPJB}, we do all calculations in the TB approximations
with the nearest-neighbor interactions only.
If not differently specified, all the hopping matrix elements have the value
of $\tB$.
We will compare the results of our generalized SWPA
with the results of Ref.~\onlinecite{our_EPJB}.
%
%
\subsection{\label{ssec:stat_curr}Stationary currents: convergence with the basis set size}
%
%
First we calculate stationary currents for the 16-atoms ring, as it has been done
using the purely stroboscopic basis set in Ref.~\onlinecite{our_EPJB}.
%
\begin{figure}[!thb]
\centerline{\includegraphics[width=86mm]{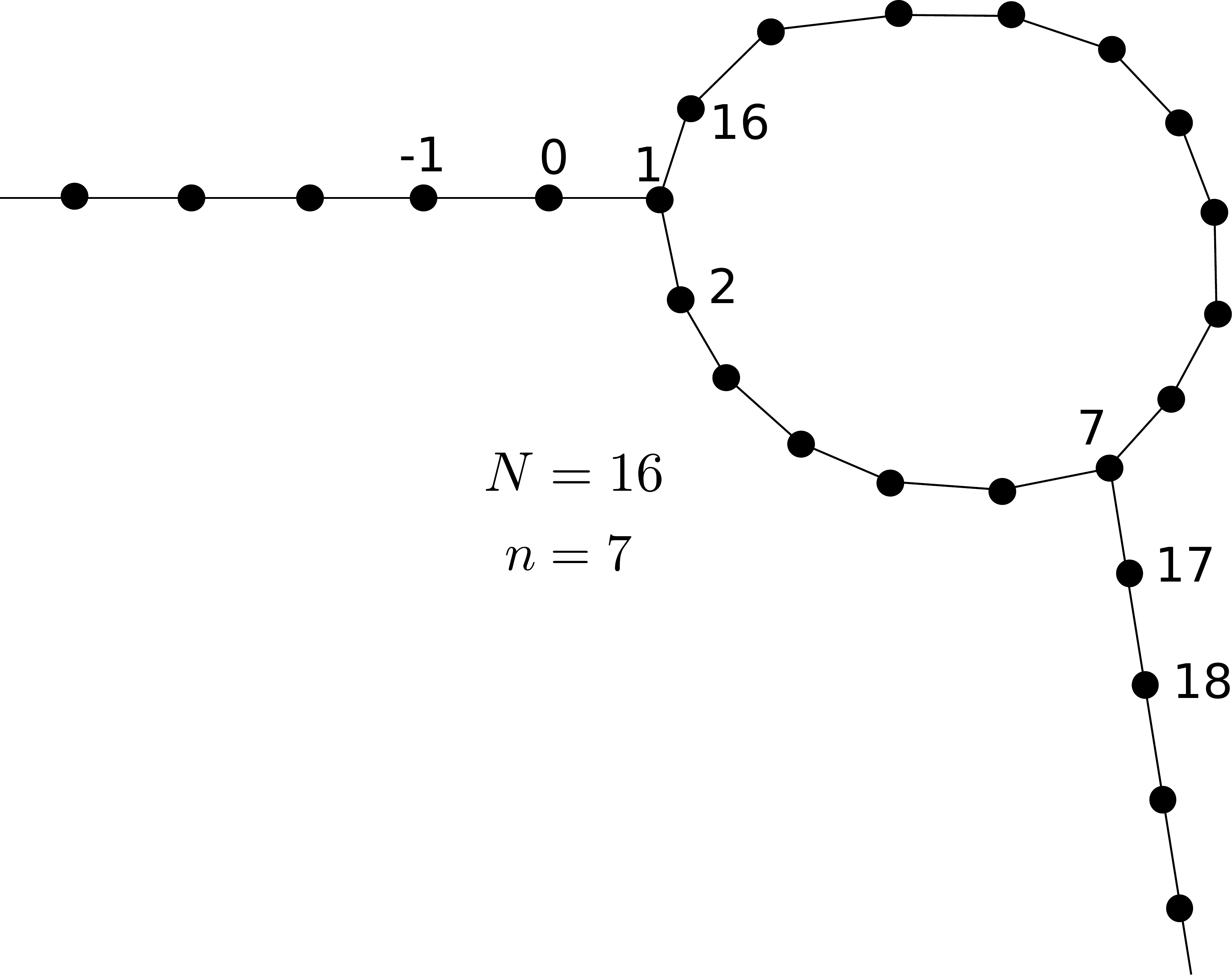}}
\caption{Topology of the ring with the two electrodes attached to its vertex atoms.
This system has also been analyzed by our former time-dependent
method~\cite{our_EPJB}.
$N$ has the meaning of the total number of the ring atoms
(as well as for the index of the atom just above the left vertex).
The index $n$ denotes the index of the right-vertex atom.
The indexing scheme in this figure is the same as was used in our older work~\cite{our_EPJB}.}
\label{fig:ring_struct}
\end{figure}
%
The ring has its vertex atoms at sites 1 and 7 (see Fig.~\ref{fig:ring_struct}).
The current is calculated in between the sites $120$ and $121$ 
(in the drain electrode) in the indexing scheme of our older work.
We show and compare the results on Fig.~\ref{fig:ring_stat_curr}.
As we can observe, our present method with the mixed
stroboscopic+atomic orbital basis set practically matches the stationary results from exact eigenstates
(which are shown for reference and were computed in Ref.~\onlinecite{our_EPJB}).
This agreement is obtained already at a very moderate value of $\mmax = 50$.
In contrast,
employing the purely stroboscopic basis set, we were not able to obtain converged
results at any affordable $\mmax$ value.
The small departure of the present results from the exact ones occurs only
at very high bias values close to $2\,|\tB|/e$.
Even this imperfection can easily be corrected using a slightly higher $\mmax$.
This is demonstrated by the plot in the inset which uses the fixed value of the bias.
%
\begin{figure}[!thb]
\centerline{\includegraphics[width=86mm]{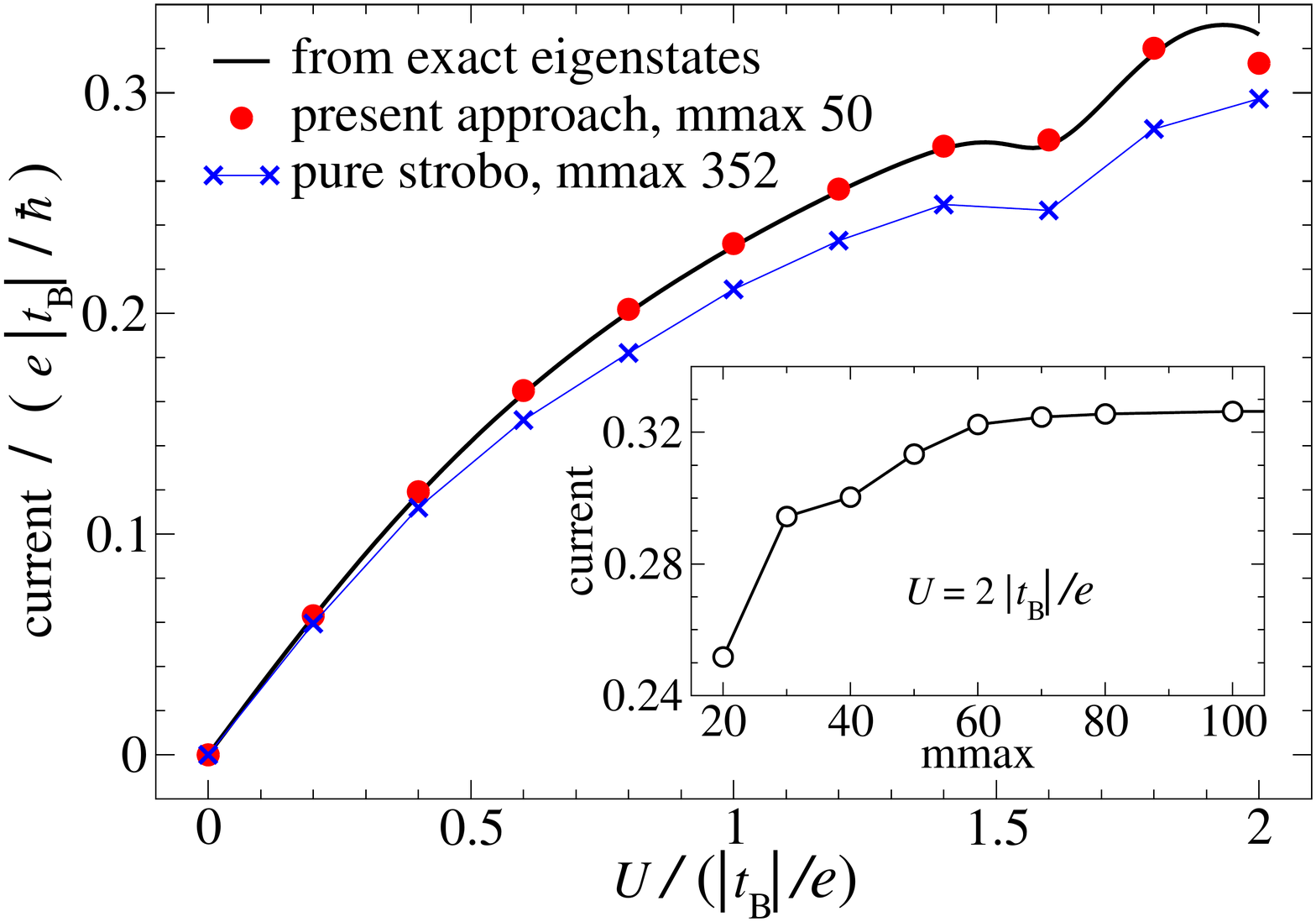}}
\caption{Stationary electron currents through the leads of
the system depicted on Fig.~\ref{fig:ring_struct} plotted as functions of the bias voltage $U$.
The results have been calculated by the three different approaches
for the purpose of the comparison.
Solid black line shows the exact stationary electron currents obtained
in Ref.~\onlinecite{our_EPJB}.
Discrete symbols show long-time quasi-stationary values from our
time-dependent approaches:
The blue x crosses show results adapted from our work~\cite{our_EPJB}
which have been obtained using the the purely stroboscopic basis set
with the (relatively large) cutoff given by $\mmax = 352$.
The full red circles show results from our present approach which employs 
the mixed basis set corresponding to $\mmax=50$,
$M_0 \equiv N = 16$ (the ring size) and $M_1 = M_2 = 2$.
The inset show results calculated by the present method at bias $U = 2$ units
for several parameters $\mmax$ (to which the basis set size is directly
proportional).
More details can be found in the text.}
\label{fig:ring_stat_curr}
\end{figure}
%
We recall that all the quasi-stationary currents displayed
in Fig.~\ref{fig:ring_stat_curr} have been computed in the bond between
sites $120$ and $121$, i.e. in the drain electrode.
Choosing a site closer to the ring would fully correct the convergence issue
even for the single unconverged result at $U = 2\,|\tB|/e$.
(Using a limited number of the stroboscopic wave packets, only a finite
portion of the real space around the lattice center is covered by these
partially localized basis functions.
This is the origin of the single unconverged result
in Fig.~\ref{fig:ring_stat_curr}.)
On the contrary, in the original SWPA~\cite{our_EPJB} we could not obtain
full convergence even at sites very close to the lattice center.
Otherwise, within the well covered region, the computed quasi-stationary
currents are practically independent on chosen position in the electrodes,
except of the rapid unphysical oscillations in time-dependent currents
(see below) which could still be suppressed using a higher $\mmax$.

The substantial improvement of the convergence in the present method in comparison
to the original SWPA arises from the employment of
localized basis functions for the central system (the ring with some finite pieces of the electrodes)
and from the employment of the independent subsets of the stroboscopic wavepackets, one subset per each lead.
%
%
\subsection{\label{ssec:circulating} Time-dependent circulating currents}
%
%
Having viewed the stationary currents in the previous subsection, 
we now proceed with the time dependent regime for which the SWPA has been primarily developed.
We choose the ring system with $N \equiv M_0 = 18$ atoms and with the right electrode attached to atom $n=5$.
The indexing used for presentation of our results has been explained in Fig.~\ref{fig:ring_struct}.
The left electrode is always being attached to the atom with index $1$. 
For this system it was particularly difficult to obtain converged results in the past~\cite{our_EPJB}.
Regarding the physical effects, it is an interesting structure as it exhibits
significant circulating currents for a wide range of biases~\cite{Stefanucci09}.
In Ref.~\onlinecite{our_EPJB} we have discussed this effect also in the time domain.
In this subsection we present how our generalized SWPA suits for treatments of the reported effects.
%
\begin{figure}[t]
\centerline{\includegraphics[width=86mm]{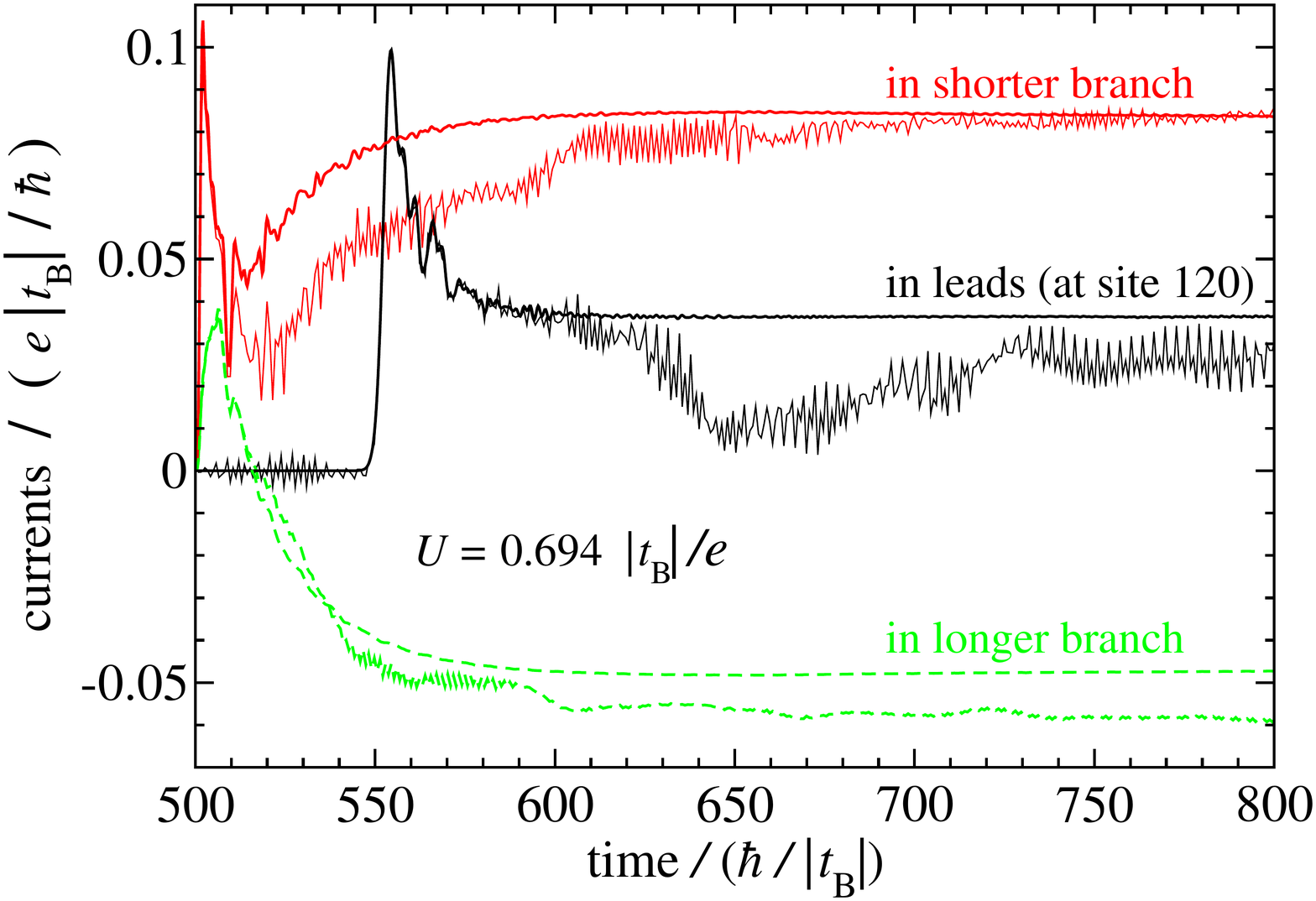}}
\caption{Time-dependent electron currents in the drain electrode and in the two
ring branches, computed for the $N=18$, $n=5$ ring.
The black (dark) solid lines represent the currents through the electrodes,
calculated between the sites 120 and 121.
The red (gray) solid lines represent the currents through the shorter ring branch.
The green (light gray) dashed lines show values of the current in the longer ring branch.
The plots with the rapid oscillations and/or noisy quasi-stationary
curves have been obtained using the original stroboscopic method
with $\mmax=1200$.
New results (the smoother plots) have been obtained using the mixed
stroboscopic+atomic orbital basis set with $\mmax=50$ and $M_1 = M_2 = 150$.
In this figure we do not use any running averages of the non-physical rapid oscillations.}
\label{fig:ring_circul_curr}
\end{figure}
%
The best picture is obtained by direct comparison to the older results from 
the original stroboscopic method~\cite{our_EPJB}.
The system is first equilibrated by 
evolving it in time without any bias for a period of $500$ units of time.
At time $t_\mathrm{sw} = 500\,\hbar/|\tB|$ we abruptly turn on the bias voltage
$U$, which remains constant afterwards.
The particular value $U = 0.694\,|\tB|/e$ was chosen in order to maximize
the effect of the circulating current~\cite{our_EPJB}.
Resulting time-dependent electron currents are shown in Fig.~\ref{fig:ring_circul_curr}.
The currents in the branches have been obtained by taking averages from local currents in particular bonds
of given branch.
The plots with the rapid oscillations (or noisier quasi-stationary dependencies at larger times)
have been obtained by our original SWPA.
The much smoother plots come from the present generalized SWPA.\footnote{We remark that
ring $(N,n) = (18,5)$ was quite basis-set demanding within our older approach.
The basis-set effect has revealed itself also at the intermediate times,
around $650\,\hbar/|\tB|$ in Fig.~\ref{fig:ring_circul_curr} -- see the drop of the
black oscillating plot.}
In contrast to Ref.~\onlinecite{our_EPJB} we present all results
(including the old ones)
without any smoothening of the non-physical rapid oscillations which are
prominent in the older results.
While the older results have employed a very large basis set corresponding
to $\mmax=1200$, now we use $\mmax=50$ only and in addition we use
a relatively large number of the localized atomic orbitals: $M_1 = M_2 = 150$
which together with $N \equiv M_0 = 18$
sum up to $318$ localized atomic orbitals covering the formal device.
The reason behind 
the relatively large number of electrodes' atoms included into the formal device
is that we decided to compute the local current
at the relatively distant bond from the ring (about 120 lattice parameters).
Using the localized atomic orbitals as basis functions helps to obtain smoother
local current
(with suppressed unphysical oscillations).\footnote{Compare it to the ring studied
in subsection~\ref{ssec:stat_curr} where we have
added only 2 atoms per lead ($M_1 = M_2 = 2$ in that case) into the formal device.
The rapid unphysical oscillations were not of concern there as we only needed
the long-time quasi-stationary currents in that case.
They were obtained by taking proper averages of the rapidly oscillating curves.}
Other way to reduce the oscillations would be to apply a much higher value
of $\mmax$.
The ``mirror'' site indices $M_1$ and $M_2$ could in such calculation be small
and the oscillations would still be reduced.
This alternative approach would be much more expensive however.

We end the present subsection by the conclusion that the generalized SWPA
is able to obtain practically converged results, in contrast to the original
approach used in Ref.~\onlinecite{our_EPJB}.
%
\begin{figure}[t]
\centerline{\includegraphics[width=0.49\textwidth]{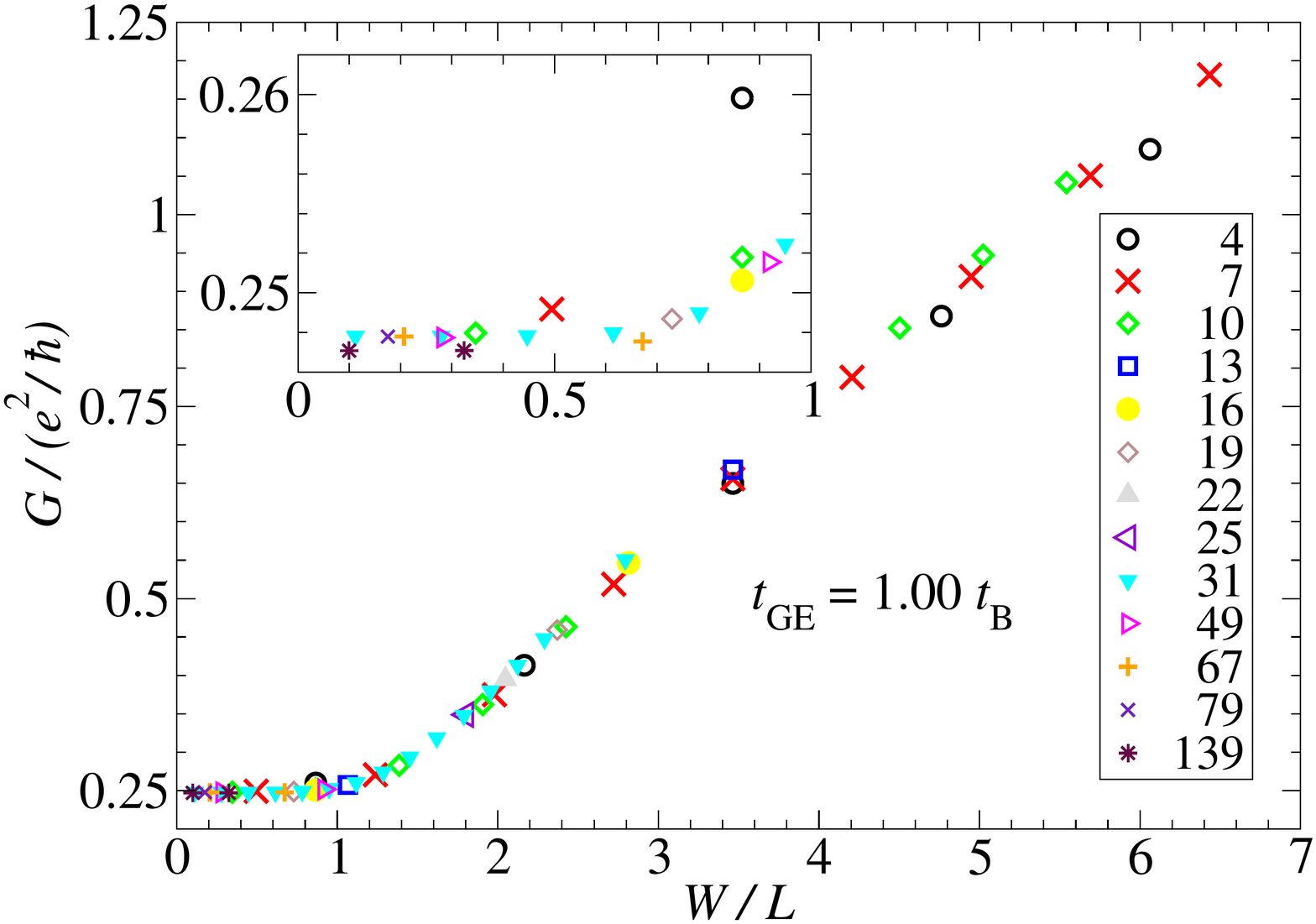}}
\caption{Conductances of various metallic AGNRs of the same symmetry as shown in Fig.~\ref{fig:GNR_img_basic}.
The AGNRs lengths $L = L_\mathrm{tot}-4a$ are shown in legends.
The conductances are plotted against the ratio $W/L$.
The coupling parameter between the GNR and the leads is $\tGE = \tB$.
The inset magnifies the interval of $W/L \le 1$.}
\label{fig:GNR_GWLrat_coup1.00}
\end{figure}
%
The contrast is even more striking when we take into account that calculation
of the plots on Fig.~\ref{fig:ring_circul_curr} employing the older method 
took about 4 weeks of machine (a 4-core 2.0 GHz Xeon server) time
and about 24 GiB of memory.
The much improved results obtained by the new method needed only 2 minutes
for the same task.
The enormous reduction of computational demands is given by the fact that
in the new method we do not need to compute and store many overlaps
$\langle L,l|L,n,m;t\rangle$ (see subsection~\ref{ssec:SchE}).
The only overlaps which are necessarily needed are those between the
stroboscopic wave packets (all of them)
and the mirror sites.
I.e. the number of necessary overlaps is equal to the number of the
stroboscopic wave packets in one lead which is $\Nb (2\mmax+1)$.
In contrast, the original method~\cite{our_EPJB} requires overlaps for many
(in practice thousands)
atomic orbital of the lead with a bias voltage applied on.
Hence the number of the necessary overlaps in the original method
scales roughly as $\Nb (2\mmax+1) l_\mathrm{max}$,
where $l_\mathrm{max}$ is the highest lattice site included
(several thousands).
%
%
\section{\label{app:WLrat}Conductance dependence on the AGNR dimensions}
%
%
Here we provide calculated dc conductances $G = \lim_{U \to 0} I/U$ for a set of AGNRs of the metallic type, all having
spatial symmetries like the AGNR shown on Fig.~\ref{fig:GNR_img_basic} and using the model of the electrodes
composed of the collections of monoatomic chains. (See main text.)
%
\begin{figure}[t]
\centerline{\includegraphics[width=0.49\textwidth]{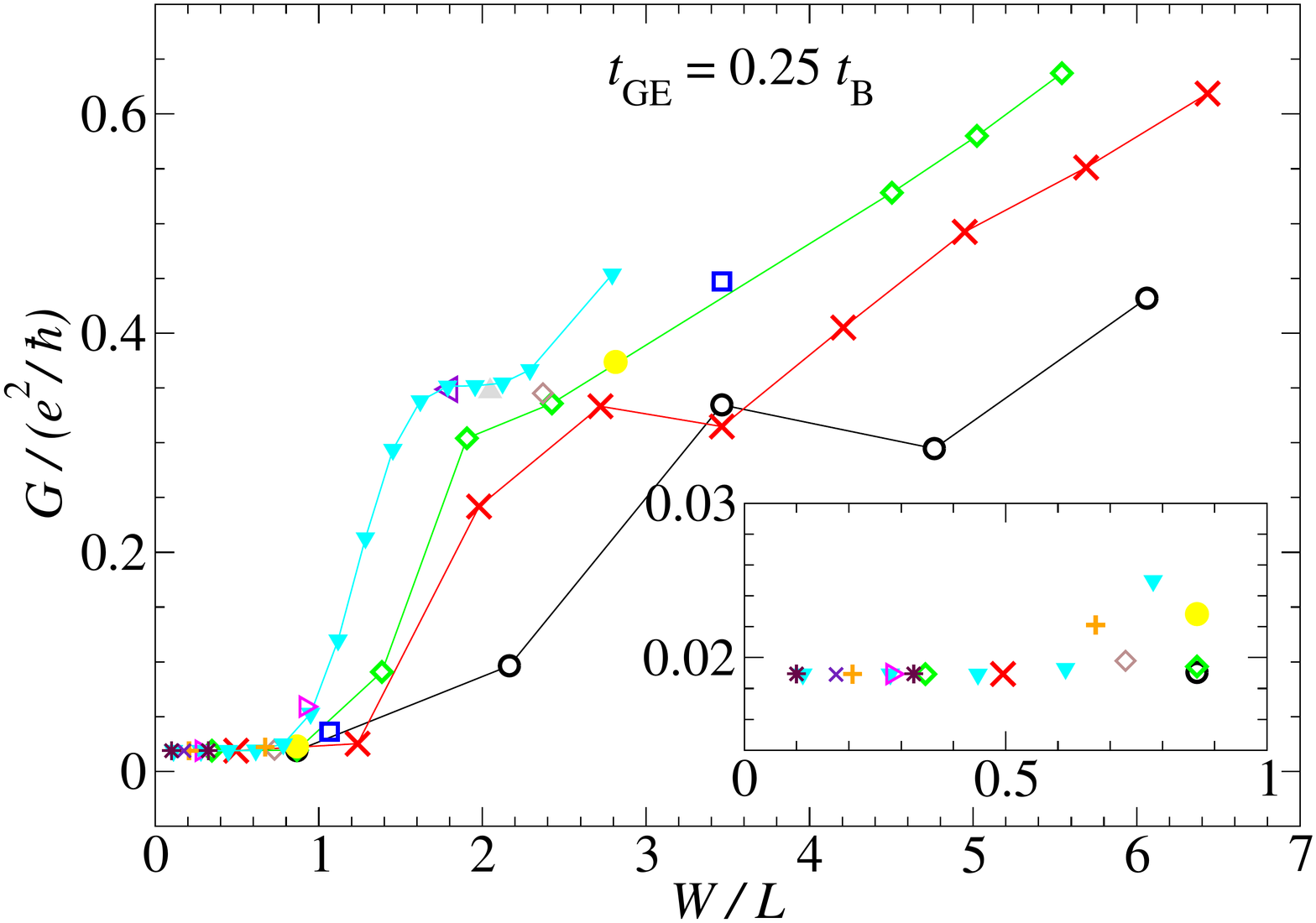}}
\caption{Same as on Fig.~\ref{fig:GNR_GWLrat_coup1.00} but now for the reduced coupling between the GNR and the leads:
$\tGE = 0.25\,\tB$.
The key to point symbols is the same as in Fig.~\ref{fig:GNR_GWLrat_coup1.00}.
The lines connecting the point symbols are guides to an eye and they are used only in cases when more than 2 widths
were considered for particular length.}
\label{fig:GNR_GWLrat_coup0.25}
\end{figure}
%
Calculations were performed using the same stationary formalism as employed to obtain the line plots in
Fig.~\ref{fig:GNR_Itdep_Biases}c.
The results on the figures are provided for the limited set of nano-ribbons ($52$ AGNRs in total).
For some of the lengths $L$ we have considered only 1 or 2 widths $W$.
The narrow-AGNR limit ($W/L \ll 1$) in Fig.~\ref{fig:GNR_GWLrat_coup1.00} corresponds to the conductivity
$G_\mathrm{narr} = 0.779\, (2e^2/h)$.
The value of the transmission coefficient, $T_\mathrm{narr} = 0.779$, is specific for our model of the electrodes
(see sec.~\ref{ssec:model}).
For electrodes considered as continuations of the AGNR, the transmission coefficient would be $1$,
see Refs.~\onlinecite{Katsnelson06,Beenakker06,White07,Onipko08,Mucciolo09,Perfetto10}.
In the opposite limit ($W/L \gg 1$) in which $G$ is no more a constant (but $\sigma$ becomes a one) we obtain 
$\sigma_\mathrm{wide} = 0.908\, (4/\pi) \, (e^2/h)$ which is close to the well-known graphene minimum dc conductivity
$\sigma_\mathrm{dc} = (4/\pi) \, (e^2/h)$.
Again, the difference of the prefactor $0.908$ from the unity is due to our specific model of the electrodes.
In case of the reduced AGNR-electrodes coupling (Fig.~\ref{fig:GNR_GWLrat_coup0.25}) the situation is much more
complicated.
Still there is a universal value of the conductance in the $W/L \ll 1$ limit (in the sense that it does not
depend on neither $W$ nor $L$ within this limit).

\begin{thebibliography}{99}
\bibitem{reduced_basis}
The reduced basis set size
used for panel (c) of Fig.~\ref{fig:GNR_Itdep_Biases} (the isolated x symbols)
practically did not affect the long-time values,
only introduced a numerical noise which was not relevant for reading out the long-time values
(obtained with the aid of proper averaging over the noisy data).
\bibitem{symmetry}
We remark that for a meaningful study of the $G(W,L)$ scaling one has to compare GNRs of the same symmetry
and of the same conductivity type (here metallic).
For wide AGNRs ($W/a \gg 1$), the condition on the metallicity is not important as in this regime all AGNRs
are conductive.
\end{thebibliography}
\end{document}